\documentclass[12pt]{article}

\usepackage{hyperref}

\newcommand{\be}{\begin{equation}}
\newcommand{\ee}{\end{equation}}
\usepackage{amsmath}
\usepackage{amssymb}
\usepackage{latexsym}
\usepackage{epsfig}

\newcommand{\ord}[1]{\mathcal{O}(#1)}
\newcommand{\bra}[1]{\langle #1|}
\newcommand{\ket}[1]{|#1\rangle}
\newcommand{\braket}[2]{\langle #1|#2\rangle}

\newcommand{\BRa}[1]{\Bigl\langle #1\Bigr|}
\newcommand{\KEt}[1]{\Bigl|#1\Bigr\rangle}

\newcommand{\BB}{{\cal B}}
\newcommand{\CC}{{\cal C}}
\newcommand{\DD}{{\cal D}}
\newcommand{\OO}{{\cal O}}
\newcommand{\LL}{{\cal L}}
\newcommand{\PP}{{\cal P}}
\newcommand{\refb}[1]{(\ref{#1})}
\newcommand{\wt}{\widetilde}
\newcommand{\RRR}{{\hbox{\rm R\kern-2.35mm R}}}
\newcommand{\p}{\partial}

\def\ZZZ{{\hbox{ Z\kern-1.6mm Z}}}

\newcommand{\RR}{{\cal R}}

\newcommand{\BQBs}[2]{\frac{{\cal B}_{(#1)}}{{\cal L}_{(#1)}}
\, Q \, \frac{{\cal B}^\star_{(#2)}}{{\cal L}^\star_{(#2)}}}
\newcommand{\BsQB}[2]{\frac{{\cal B}^\star_{(#1)}}{{\cal L}^\star_{(#1)}}
\, Q \, \frac{{\cal B}_{(#2)}}{{\cal L}_{(#2)}}}
\newcommand{\BQB}[2]{\frac{{\cal B}_{(#1)}}{{\cal L}_{(#1)}} \, Q \,
\frac{{\cal B}_{(#2)}}{{\cal L}_{(#2)}}}
\newcommand{\BsQBs}[2]{\frac{{\cal B}^\star_{(#1)}}{{\cal L}^\star_{(#1)}}
\, Q \, \frac{{\cal B}^\star_{(#2)}}{{\cal L}^\star_{(#2)}}}

\begin{document}

\begin{titlepage}
\rightline{\tt arXiv:0712.0627}
\rightline{\tt MIT-CTP-3917}
\begin{center}
\vskip 1.5cm
{\Large \bf {Linear b-Gauges for Open String Fields}}\\

\vskip 1.5cm
{\large {Michael Kiermaier${}^1$,
Ashoke Sen${}^2$, and Barton Zwiebach${}^1$}}
\vskip 1.0cm
{\it {${}^1$ Center for Theoretical Physics}}\\
{\it {Massachusetts Institute of Technology}}\\
{\it {Cambridge, MA 02139, USA}}\\
mkiermai@mit.edu, zwiebach@.mit.edu
\vskip .5cm
{\it {${}^2$ Harish-Chandra Research Institute}}\\
{\it Chhatnag Road, Jhusi, Allahabad 211019, INDIA}\\
sen@mri.ernet.in

\vskip 2.0cm
{\bf Abstract}

\vskip 1.0cm

\end{center}

\noindent
\begin{narrower}

Motivated by Schnabl's gauge choice, we explore
open string perturbation theory in 
gauges where a linear combination
of antighost oscillators annihilates the string field.
We find that in these
linear $b$-gauges
different gauge conditions are needed
at different ghost numbers.
We derive the full propagator
and prove the
formal properties
which guarantee that the Feynman diagrams reproduce the
correct on-shell amplitudes.
We find that these properties can fail due to the need
to regularize the propagator, and identify
a large class of  linear $b$-gauges
for which they hold rigorously.
In these gauges  the propagator has a
non-anomalous Schwinger representation
and builds Riemann
surfaces by adding strip-like domains.
Projector-based gauges, like Schnabl's,
 are not in this class of gauges
but we construct
a family of regular linear $b$-gauges which interpolate
between Siegel gauge and Schnabl gauge.

\end{narrower}

\medskip

\end{titlepage}

\newpage

\baselineskip=16pt

\tableofcontents

\section{Introduction and summary} \label{s1}

Since the discovery of interacting open bosonic
string field theory~\cite{wit1}, much
work has been devoted to understanding the Feynman rules of the
theory and deriving the Polyakov amplitudes using these Feynman
rules~\cite{gid,gidmar,boch,thorn}.
Most of these studies have
been carried out in the Siegel
gauge~\cite{siegel1}.
 More recently,
Schnabl's discovery~\cite{0511286} of an
 analytic classical solution in string field theory in a different
 gauge has inspired a large amount of work on open string field theory
 in \emph{Schnabl gauge} and closely related gauges.
Most of these studies
focus on finding classical solutions of
open string field theory and/or studying various properties
of these
solutions~\cite{0603159, 0603195, 0605254, 0606131, 0606142,
0610298, 0611110, 0611200, 0612050, 0701248, 0701249, 0704.0930,
0704.0936, 0704.2222, 0704.3612, 0705.0013, 0706.0717,
0707.4472, 0707.4591, 0708.3394, 0709.2888, 0710.1342}.

There has also been some
progress towards obtaining
the Feynman rules of string field
theory in this new class of
gauges and computing
off-shell
amplitudes~\cite{0609047, 0708.2591}
(see also~\cite{0611189}).
The off-shell Veneziano amplitude in Schnabl gauge was obtained
in~\cite{ 0708.2591}, completing the work
of~\cite{0609047}.
There was a surprise.  The amplitude
receives contributions from terms whose
Siegel gauge analogs would vanish.
These contributions require delicate
regularization of the propagator
which makes the construction of general tree
amplitudes quite nontrivial.
Motivated by this puzzle in this paper we
carry out a systematic study of string perturbation theory in
a wide class of gauges which we shall call `linear $b$-gauges'.
These gauges include both Siegel gauge and Schnabl gauge as special cases.
Other special cases of such gauges have been studied
previously in~\cite{preit}.

Our analysis demystifies some of the results found
in~\cite{0708.2591}.
We find that the  delicate contributions which arise at tree level
occur because the propagator fails to move the open string midpoint.
Moreover, we show
that what has so far been called the Schnabl-gauge
propagator is  the correct propagator only at the
string tree level.
For string loop diagrams we need to include string fields of
all ghost numbers~\cite{boch,thorn}, and
the propagator takes  different form   in different ghost-number sectors.
This is a general
feature of linear $b$-gauges.
Even after taking this effect into account the
proof of consistency of Feynman
amplitudes  is complicated in
Schnabl gauge, again because the propagator
does not move the open string midpoint.
Motivated by this observation we derive  a
set of conditions which guarantee
that a linear $b$-gauge defines a consistent perturbation theory.
 This is one of our main results.
Schnabl gauge fails to satisfy these
conditions.\footnote{Since our conditions are
sufficient but not necessary for a gauge choice to be valid, the failure
of the Schnabl gauge to satisfy our condition does not immediately
rule it out as a valid gauge choice.  It shows, however, that
establishing consistency of string perturbation theory in Schnabl gauge
is a much more difficult task.}
During the course of our analysis
we obtain an interesting and explicit
Riemann surface interpretation of the propagator
for general linear $b$-gauges.
We also construct a family of
 regular 
linear $b$-gauges
which interpolate between
 Schnabl gauge and  Siegel gauge.

\medskip

We shall now summarize the main results of the paper.
As is well known, a general quantum string field $|\psi\rangle$
is described
by a state in the first quantized open string state space with
arbitrary ghost number. After suitable gauge fixing
the action takes the form:
\be \label{eas0}
S = -\left[{1\over 2} \bra\psi Q \ket\psi + {g_o\over 3}
\bra\psi \psi * \psi\rangle\right]\, .
\ee
Here $Q$ denotes the BRST operator, $*$ denotes star product,
and $g_o$ is the open string coupling constant. This form
of the action may be obtained either by starting with the classical
string field theory (which has the same action but $\ket\psi$
restricted to ghost-number one) and going through the Fadeev-Popov
procedure or by using the Batalin-Vilkovisky formalism.
We choose the gauge condition on the
ghost-number $g$  string
field $\ket{\psi_{(g)}}$ as
\be \label{eas1}
\BB_{(g)}\ket{\psi_{(g)}} = 0\, ,
\ee
where $\BB_{(g)}$ is a linear combination
of the oscillators $b_n$ that can be encoded in a vector
field~$v(\xi)$:\footnote{
At this point we regard the vector field $v(\xi)$ as a
formal Laurent series in $\xi$.
 Later we will
 demand that this Laurent series defines an
 analytic function in some neighborhood
 of the unit circle $|\xi|=1$. }
\be \label{evector}
\BB_{(g)}\equiv \sum_{n\in \mathbb{Z}}
v_n b_n = \oint{d\xi\over 2\pi  i}
\,
v(\xi) b(\xi)
\, , \qquad\hbox{with} \quad
v(\xi) = \sum_{n\in \mathbb{Z}} v_n \xi^{n+1}\, .
\ee
For each ghost-number $g$ we need a vector field to define the
operator  $\BB_{(g)}$.
We find that the consistency of gauge fixing requires us to choose
\be \label{eas2}
\boxed{ \phantom{\Bigl(} \,\,\BB_{(3-g)} = \BB_{(g)}^\star\, ,~}
\ee
where $\BB_{(g)}^\star$ denotes the BPZ
conjugate of $\BB_{(g)}$. We shall refer to this as
a linear $b$-gauge.
Siegel gauge corresponds to the choice $\BB_{(g)}=b_0$ for
all $g$. In Schnabl gauge  we have
\be \label{new15}
\BB_{(1)}
= B \equiv
b_0 +
2\sum_{k=1}^\infty {(-1)^{k+1}\over 4 k^2 -1}
 \, b_{2k}\,, \quad   v(\xi) =  (1+ \xi^2) \tan^{-1} \xi  \,.
\ee
For a general linear $b$-gauge  $\BB_{(g)}$ is
not invariant under BPZ conjugation and eq.\refb{eas2} prevents
us from choosing the same gauge condition on all ghost
sectors.  In particular, Schnabl's $\BB_{(1)}$ is not BPZ
invariant
and
{\em cannot} be used for all ghost numbers.  One must have
$\BB_{(2)} = \BB_{(1)}^\star \not= \BB_{(1)}$.
A natural possibility consistent with \refb{eas2}  is to take
  \be \label{eas15}
\BB_{(g)} = \begin{cases}{ B \quad \, \,
\hbox{for $g$ odd,}} \cr
{B^\star \quad \hbox{for $g$ even.}}\end{cases}\,
\ee
Both Siegel gauge and Schnabl gauge
are examples in which $\BB_{(g)}$, for a
given $g$,  is a
linear combination of  $b_n$ modes with $n\ge 0$  or with
$n\leq 0$.
We will  also
be able to handle the case of linear
combinations of $b_n$ modes
with  both positive and negative $n$.

It is useful to assemble all the $\BB_{(g)}$ operators  into
a single operator $\BB$ defined by
\be \label{eps1}
\BB =\sum_g \BB_{(g)}\Pi_g\, ,
\ee
where $\Pi_g$ is the  projector
onto ghost-number $g$ states.
Acting on a ghost-number $g$ state
we have
$\BB=\BB_{(g)}$, and the
gauge-fixing condition \refb{eas1}
becomes
\be \label{eps2}
\BB \ket\psi = 0\, .
\ee

There are some possible subtleties
in defining and manipulating the propagators
in a general linear $b$-gauge, just like in the case of
Schnabl gauge~\cite{0708.2591}.
We shall first ignore these subtleties and summarize our results
in formal terms and then describe how we address these
subtleties.
In order to calculate the propagator
in the gauge~(\ref{eps2})
we introduce
ghost-number $g$ sources $\ket{J_{(g)}}$, add to the free
string field theory action
$-{1\over 2} \sum_g \bra{\psi_{(g)}} Q \ket{\psi_{(2-g)}}$
the source term $\sum_g\bra{\psi_{(g)}}J_{(3-g)} \rangle$,
and
eliminate $\ket{\psi_{(g)}}$ by its linearized
equation of motion
in the gauge \refb{eas1}. The result  is
${1\over 2} \sum_g
\bra{J_{(4-g)}} \PP\ket {J_{(g)}}$, with the full propagator $\PP$
given by
\begin{equation}\label{eas3xxx}
\boxed{ \phantom{\Biggl( }
\PP =\sum_g \PP_{(g)}
\Pi_g\, ,\quad\hbox{with} \quad
    {\cal P}_{(g)}=\frac{{\cal B}_{(g-1)}}
    {{\cal L}_{(g-1)}}
\, Q \, \frac{{\cal B}_{(g)}}{{\cal L}_{(g)}}\,,
\quad \hbox{and}\quad
 {\cal L}_{(g)} \equiv  \{ Q, {\cal B}_{(g)}\} \,.~}
\end{equation}
Note that at each ghost number the propagator involves the
gauge-fixing  operators  $\BB_{(g)}$
of two ghost numbers.  Using \refb{eas3xxx}
one can prove the fundamental property
\be \label{eas5}
\boxed{\phantom{\Bigl(} ~\{ Q, \PP\} = 1\, .~}
\ee
We will show that eq.\refb{eas5}
guarantees the decoupling of trivial states
from on-shell scattering amplitudes.
Moreover, it ensures that the
$b$-gauge propagator $\cal{P}$
gives the same on-shell amplitudes as the familiar
Siegel gauge propagator
$\overline{\cal P}= b_0/L_0$. The steps which
lead to this conclusion
are straightforward.
Since we also have $\{ Q , \overline \PP\} =1$,
it follows that
the difference of propagators
$\Delta \PP = \PP - \overline\PP$ is annihilated
by $Q$, \i.e.\ $\{ Q ,\Delta\PP\}=0$.
We find
that $\Delta \PP$ is in fact
a BRST
 commutator:
 \begin{equation}\label{eqdelp}
     \Delta \PP=[Q,\Omega ]\,,
 \end{equation}
 for some operator $\Omega$.
 As a result, given any
 amplitude in the linear $b$-gauge, we can replace
 each propagator $\PP$  by
$\overline{\cal P} + [Q,\Omega ]$.
We show that the contribution from the
$[Q,\Omega ]$ piece vanishes
 for on-shell amplitudes
after summing over Feynman diagrams.
The proof involves the same kind of cancelations
which prove that pure-gauge states decouple
from on-shell amplitudes  in Siegel gauge.
The combinatoric factors are somewhat different
but
they work out correctly.

\medskip

As anticipated above, not all linear
$b$-gauges are
 consistent
gauge choices.
To begin our analysis we make the natural assumption
that string field theory Feynman diagrams must have a representation
as correlators on Riemann surfaces. The propagator will not
permit this representation unless the operators ${\cal L}_{(g)}$
generate conformal transformations of open string theory. This implies
that
the vector field
 $v(\xi)$  associated with ${\cal B}_{(g)}$ must satisfy
 $\overline{v(\xi)}=v(\bar\xi)$.
We will also see that compatibility with the reality condition
on the string field requires  vector fields
$v(\xi)$ which are even or odd under $\xi\to -\xi$.
Thus a gauge choice which allows real string fields
in the gauge slice and which permits a geometric
interpretation of ${\cal L}_{(g)}$ requires
\begin{equation}\label{2conditions}
   \overline{v(\xi)}=v(\bar\xi)\,, \qquad v(-\xi)=\pm v(\xi)\,.
\end{equation}

A  rigorous proof of \refb{eas5} gives
further constraints.
The main obstruction comes from the subtleties in
defining the operators
$1/\LL_{(g)}$ which appear in the
propagator~\refb{eas3xxx}.
We of course do not expect $\LL_{(g)}$ to be invertible in
the full space of open string states. First of all it has
zero eigenvalues when acting on on-shell states
-- representatives of
BRST cohomology which satisfy the gauge condition. It may
also have additional zeroes acting on BRST trivial states
satisfying the gauge condition if there are residual gauge
symmetries. These are familiar situations which occur even
in conventional field theories, and give rise to poles in the
propagator at special values of the momentum. What one
requires is that $\LL_{(g)}$ should have a well defined inverse
acting on states of generic momentum. In particular
if we restrict the momentum to the deep Euclidean region
(more precisely in the region $k^2>1$ so that we avoid the
tachyon pole) then the inverse of $\LL_{(g)}$ should be
unambiguously defined.

In open string perturbation
theory the operator $1/\LL_{(g)}$
appears in the calculation of the string amplitudes.
 In all linear $b$-gauges that we consider,
 amplitudes have  a geometric interpretation
 in terms of Riemann surfaces. We shall see that
the requirement on $\LL_{(g)}$ described in the
previous paragraph is equivalent
to demanding that  $\LL_{(g)}$ can be inverted
 up to terms which represent Riemann surfaces with an
 open string degeneration, \i.e.\ surfaces
 localized at the boundary of the
 moduli space. Such surfaces  contain
a strip domain of infinite length. Their
contribution vanishes when the momentum flowing along the strip
satisfies $k^2>1$ because this ensures that only
positive conformal weight states propagate along the
infinitely long strip.\footnote{
This cannot be done for loop amplitudes where we need to integrate
over the internal momentum and we get non-vanishing
contributions from the tachyon and massless
states propagating in the loop.
Only after ignoring these infrared problems, which
have a well-defined physical origin,   the contributions
 from degenerate Riemann surfaces can be  dropped.
\label{foot2}}
 In summary, when we demand that for consistent gauge choices
 $1/\LL_{(g)}$ is well defined and eq.(\ref{eas5}) is satisfied, we only demand
 this to hold up to terms whose associated
 Riemann surfaces are \emph{localized at the
 boundary of the
 moduli space.}

To  illustrate this consider first the case of
Siegel gauge where the corresponding operator is
$1/L_0$. There we define $1/L_0$ as
 \be \label{eas12}
{1\over L_0} \equiv \lim_{\Lambda_0\to\infty}
\int _0^{\Lambda_0} ds\,  e^{-s L_0}\, .
\ee
A short calculation shows that we have
$L_0 \int _0^{\Lambda_0}
ds\,  e^{-s L_0} = 1 - e^{-\Lambda_0 L_0}$
 for finite $\Lambda_0$.
In a given line of a Feynman diagram
the operator
$e^{-\Lambda_0 L_0}$  inserts a
long strip of width $\pi$ and length
$\Lambda_0$ into the Riemann surface
associated with the amplitude.
In the $\Lambda_0\to\infty$ limit
we get a Riemann surface
at the boundary of the moduli space
and its contribution can be safely ignored in the sense described  above.
Thus the relation
$L_0 \int _0^{\Lambda_0} ds\,
 e^{-s L_0} = 1$  becomes exact
in the $\Lambda_0\to\infty$ limit, leading to
the definition \refb{eas12} of $1/L_0$.
The analysis for a linear $b$-gauge is similar.
We attempt to define $1/\LL_{(g)}$
for each
ghost-number $g$ as
\be \label{eas10}
{1 \over \LL_{(g)}} \equiv \lim_{\Lambda_{(g)}\to\infty}
\int_0^{\Lambda_{(g)}} ds
\,  e^{-s \LL_{(g)}}\, .
\ee
For finite $\Lambda_{(g)}$ we have
$\LL_{(g)} \int_0^{\Lambda_{(g)}} ds
\,  e^{-s \LL_{(g)}} = 1 - e^{-\Lambda_{(g)} \LL_{(g)}}$.
It turns out that
unless the operators $\LL_{(g)}$ satisfy certain conditions,
the $e^{-\Lambda_{(g)} \LL_{(g)}}$ factor
can generate contributions {\em away} from
 the boundary of the moduli space
 even in the $\Lambda_{(g)}\to\infty$ limit
and may not be ignored.
 In this case the Schwinger parametrization~(\ref{eas10}) is
 anomalous and does not provide a
 proper inverse to the operator $\LL_{(g)}$.
   Contributions to amplitudes which involve factors
 of $e^{-\Lambda_{(g)} \LL_{(g)}}$
vanish in the limit $\Lambda_{(g)}\to\infty$
if the vector field $v(\xi)$
is analytic in some neighborhood of the unit circle $|\xi|=1$ and
satisfies
\be \label{ealt2old}
v_\perp (\xi) \equiv \Re\bigl(\bar\xi v(\xi)\bigr)
> 0 \quad
\hbox{for $|\xi|=1$} \, .
\ee
As the notation indicates, $v_\perp (\xi)$ is the component of
$v(\xi)$ along the radial outgoing direction.  The above condition
states that on the unit circle $v(\xi)$ never vanishes and always
points outward.

\begin{figure}
\centerline{\epsfig{figure=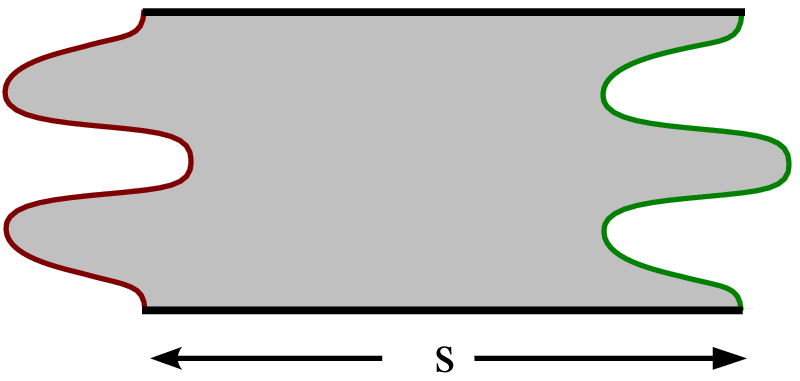, height=3cm}}
\caption{The shape of the strip $\RR(s)$ associated with the
operator $e^{-s \LL_{(g)}}$.}
\label{intro1}
\end{figure}

So far we have found that an operator ${\cal B}_{(g)}$
defines a sensible gauge condition if the associated vector
field $v(\xi)$ satisfies \refb{2conditions} and~(\ref{ealt2old}).
It is easy to see that if $v(-\xi) = v(\xi)$
we cannot satisfy \refb{ealt2old} both at $\xi$ and $-\bar\xi$.
We must require  $v(\xi)$ to be odd under $\xi\to -\xi$.
We can then summarize the conditions
on $v(\xi)$ which guarantee a regular gauge
as follows:
\begin{equation}\label{ealt2}
  \boxed{\phantom{\Biggl(}
v(\xi) = \sum_{k\in\mathbb{Z}} v_{2k}\,  \xi^{2k+1}  \quad
\hbox{with} ~v_{2k}\in \mathbb{R}\quad \hbox{and}\quad
  v_\perp (\xi) \equiv \Re\bigl(\bar\xi v(\xi)\bigr)
  > 0 \quad
  \hbox{for $|\xi|=1$} \, ,
  ~}
\end{equation}
 with $v(\xi)$ analytic in some neighborhood of
 the unit circle $|\xi|=1$.
 These conditions
 must be imposed on all the vectors needed
 to define the $\BB$ operator.\footnote{
 Condition~(\ref{eas2}) does not impose further
 constraints on the vector field $v(\xi)$
 because the BPZ dual vector  field $v^\star(\xi)$ satisfies
 \refb{ealt2} whenever $v$ does.}
 For $v(\xi)$ satisfying  (\ref{ealt2})
 in every ghost number sector, eqs.\refb{eas5} and
 \refb{eqdelp} hold strictly
 and lead to rigorous proofs of the decoupling of
 pure gauge states and the equality of on-shell amplitudes in
 linear $b$-gauges and Siegel gauge.
 Therefore a gauge choice that satisfies
 \refb{ealt2} will be called a \emph{regular  gauge}.

 We would like to emphasize that these conditions,
while sufficient for the gauge choice to be
 consistent, are not necessary. For example,
we may get a consistent gauge choice even if
$v(\xi)$ vanishes at some point $\xi_0$ on the unit circle
provided  the integral
$\int^\xi d\xi' / v(\xi')$ is finite along a contour passing
through $\xi_0$ (see footnote
\ref{footgood}). This integral can only be finite
 if $v(\xi)$ fails to be analytic at $\xi_0$.
 We will not consider such gauges in this paper.

\medskip

We show that whenever
condition \refb{ealt2} is satisfied, the insertion of the
operator $e^{-s\LL_{(g)}}$ (with $s>0$)  in a correlation function
function can be represented by
 a strip $\RR(s)$ of length~$s$ (see Fig.~\ref{intro1}).
The coordinate frame for this representation is naturally provided
by the Julia equation.
 The width of the strip $\RR(s)$ is non-vanishing, finite,
 and independent of $s$.
 The strip $\RR(s)$ is bounded above and below by a pair of
horizontal lines with open string boundary conditions.
Unlike the
rectangular strip associated with $e^{-s L_0}$,
the left and the right edges of $\RR(s)$,
which are glued to the
rest of the Riemann surface, are ragged.
In fact they
are identically shaped smooth curves
of finite horizontal spread.
We prove that in
the $s\to\infty$
limit the insertion of $\RR(s)$
gives  a degenerate Riemann
surface -- a surface at the boundary of the moduli space.
Using this we  show that the extra terms which arise
 in the calculation of amplitudes
due to the regularization
of $1/\LL_{(g)}$
are localized near the boundary of the moduli space and
can be~ignored.  We also explain geometrically
why amplitudes in
 linear $b$-gauges  
other than Siegel
gauge cannot exhibit off-shell factorization.  This failure
of off-shell factorization was investigated in detail in~\cite{0708.2591}
for the Veneziano amplitude in Schnabl gauge.  We note, however, that
off-shell factorization, while elegant and convenient, is not
a physical requirement of amplitudes.
\medskip

The Schnabl gauge condition~\refb{new15} does {\em not}
satisfy condition~\refb{ealt2} because the  vector field $v(\xi)$
associated with $\BB_{(1)} = B$
 vanishes at the point $\xi=i$ on the unit disk: $v(i) =0$.
We can
find a family of regular
gauge choices by taking
\be \label{eas15.1}
\BB_{(g)} = \begin{cases}{ \,~B^\lambda\, \,\,\,~
\quad \hbox{for $g$ odd,}}
\cr
{\left(B^{\lambda}\right)^\star
\quad \hbox{for $g$ even,}} \end{cases}\,
\ee
where
\be \label{eas16}
B^\lambda \equiv e^{\lambda L_0} B e^{-\lambda L_0} = b_0
+ 2\sum_{k=1}^\infty {(-1)^{k+1}\over 4 k^2 -1} \, e^{-2k\lambda}
\, b_{2k}\, ,
\qquad 0< \lambda < \infty\, .
\ee
The vector field associated with
$B^\lambda$ is  $v^\lambda(\xi)
= e^\lambda v (e^{-\lambda} \xi)$, where $v(\xi)$ is the vector
associated with $B$ (see~\refb{new15}).
For $\lambda >0$ the vector $v^\lambda$ satisfies condition \refb{ealt2}.
For $\lambda\to 0$ this gauge
approaches 
Schnabl gauge. On the other hand as $\lambda\to\infty$ this gauge
goes over to the Siegel gauge. Thus we have a family of
regular
gauges which interpolate between
Siegel gauge and
Schnabl gauge.

\section{General linear $b$-gauges} \label{s2}
\setcounter{equation}{0}

  In this section we shall describe general linear $b$-gauges
and the associated propagators.
In \S\ref{s2.1} we
explain in detail the linear
$b$-gauge conditions on the string field.
\S\ref{s2.2} will be devoted to the computation of the propagator
in a general linear $b$-gauge. In \S\ref{s2.3}
we describe
some algebraic properties of the propagator which will
be useful  in \S\ref{s3} for studying amplitudes in string field theory.
In \S\ref{sreality}
we analyze
 the conditions under which a linear $b$-gauge can be considered
 a physically reasonable gauge choice.
Finally in \S\ref{s2.4} we give some explicit
examples of linear $b$-gauges.

\subsection{Gauge conditions, ghosts, and gauge fixed action}
\label{s2.1}

The gauge-fixing procedure begins by imposing
a gauge condition on the classical open string fields,
\i.e.\ the fields $\ket{\psi_{(1)}}$ at ghost number one.
The free string field theory that includes these fields
is simply
\begin{equation}
S_1 = -{1\over 2}  \bra{\psi_{(1)}} Q \ket{\psi_{(1)}}\,.
\end{equation}
The gauge invariance $\delta_\epsilon \ket{\psi_{(1)}}
= Q \ket{\epsilon_{(0)}}$,
where $\ket{\epsilon_{(0)}}$ is an arbitrary gauge
parameter of ghost number zero,
is fixed with the gauge condition:
\begin{equation}
\label{B_introduced}
{\cal B}_{(1)} \ket{\psi_{(1)}}  = 0\, \,.
\end{equation}
The operator ${\cal B}_{(1)}$ above is some
particular linear combination
of the oscillators $b_n$.

In the Fadeev-Popov (FP) procedure one considers the gauge-fixing
functions $F_i(\psi)$, such that $F_i(\psi)=0$ are
the gauge-fixing conditions,
and writes a
FP
ghost action of the form
\begin{equation}
S_{FP} \,\sim\, \hat b^i   \,
\Bigl( \hat c^\alpha
{\delta \over \delta \epsilon^\alpha}\Bigr)
 \delta_\epsilon F_i (\psi)\,.
\end{equation}
 Here  $\hat c^\alpha$ and $\hat b^i$ are the FP ghosts
 and FP antighosts
 respectively, and $\delta_\epsilon F_i$ is
 the
 variation of the gauge fixing functions
 under infinitesimal gauge transformation with
 parameters $\epsilon^i$.
The FP antighosts
are in one-to-one correspondence with the
gauge-fixing conditions and the FP ghosts
are in one-to-one
correspondence with the gauge parameters.
Since the gauge transformation
parameters in open string field
theory are in one-to-one correspondence with ghost-number
zero states in the underlying conformal field theory (CFT),
it is natural to represent the
FP ghost fields by ghost-number zero states
 $\ket{\psi_{(0)}}$ of the CFT.
For gauge conditions of the type
\refb{B_introduced}
we can  associate the FP antighost fields
with ghost-number three states
$\bra{\tilde\psi_{(3)}}$
of the CFT, since the ghost
action may then be written as
\begin{equation}
S_2= -\bra{\tilde\psi_{(3)}} \Bigl( \psi_{(0)}
{\delta \over \delta \epsilon_{(0)}}\Bigr)
 {\cal B}_{(1)}\, \delta_\epsilon
\ket{\psi_{(1)}}=
-\bra{\tilde\psi_{(3)}}
\Bigl( \psi_{(0)} {\delta \over \delta \epsilon_{(0)}}\Bigr)
{\cal B}_{(1)}  Q  \ket{\epsilon_{(0)}}
=  -\bra{\tilde\psi_{(3)}}
{\cal B}_{(1)} Q \ket{\psi_{(0)}}\,,
\end{equation}
where the minus sign has been included for
later convenience.
It is natural to absorb the ${\cal B}_{(1)}$ factor into the
definition of the bra by setting
\begin{equation}
 \bra{\psi_{(2)}} \equiv \bra{\tilde\psi_{(3)}}
 {\cal B}_{(1)}\,,
\end{equation}
so that we have
\begin{equation}\label{first_step_gi}
S_2 =  -\bra{\psi_{(2)}}  Q \ket{\psi_{(0)}}\,.
\end{equation}
Note that $\bra{\psi_{(2)}}$ contains fewer
degrees of freedom than $\bra{\tilde\psi_{(3)}}$ since it
is subject to the condition
\begin{equation}
\label{Bonpsitwo}
\bra{\psi_{(2)}} {\cal B}_{(1)}=0   \quad \to \quad
{\cal B}_{(1)}^{\,\star} \ket{\psi_{(2)}} = 0\,.
\end{equation}
Here ${\cal B}_{(1)}^{\,\star}$
denotes the BPZ conjugate of ${\cal B}_{(1)}$.
In fact the degrees of freedom of $\bra{\psi_{(2)}}$ are in
one to one
correspondence with the gauge-fixing
conditions \refb{B_introduced} since the latter may be
expressed as $\bra s\psi_{(1)}\rangle = 0$ for arbitrary
ghost-number two states
$\bra{s}$ satisfying $\bra s \BB_{(1)}=0$.
Thus
$\bra{\psi_{(2)}}$ is a more faithful representation of the
FP antighost
fields than $\bra{\tilde\psi_{(3)}}$.
Note that the `gauge condition' \refb{Bonpsitwo}
on ghost-number two states was
preordained once we chose the gauge condition (\ref{B_introduced}) on
 states of ghost number one.

As is well known, the gauge-fixing
procedure does not stop here
since the ghost action \refb{first_step_gi}
also has gauge invariance.
This forces us
to include an infinite set of FP ghost fields represented by
CFT states of ghost number $\le 0$, and an infinite
set of FP
antighost
fields represented by CFT states of ghost
number $\ge 2$~\cite{boch,thorn}.
To proceed in a more systematic fashion, it is convenient
to introduce the full string field $\ket{\psi}$ which is
 a sum over the string fields
$\ket{\psi_{(g)}}$ of different ghost numbers $g$:
\begin{equation} \label{edeftot}
    \ket{\psi} = \sum_g\ket{\psi_{(g)}}
    \,.
\end{equation}
Let
\begin{equation} \label{exy1}
 {\cal B}_{(g)}  \ket{\psi_{(g)}}  = 0 \,,
\end{equation}
be the `gauge condition'
on $\ket{\psi_{(g)}}$.
So far  (\ref{Bonpsitwo}) tells us that
\begin{equation} \label{ebstar}
 {\cal B}_{(2)} =  {\cal B}_{(1)}^{{}^{\,\star}}\,.
\end{equation}
At the next step of gauge fixing, the gauge invariance
$\delta_\epsilon \ket{\psi_{(0)}} =
Q \ket{\epsilon_{(-1)}}$ of (\ref{first_step_gi})
requires that we impose a gauge condition
\begin{equation}\label{gaugecondPsi0}
    {\cal B}_{(0)}\ket{\psi_{(0)}}=0\,.
\end{equation}
The choice of ${\cal B}_{(0)}$ is quite arbitrary.
We need not choose
it equal to ${\cal B}_{(1)}$ or ${\cal B}_{(1)}^\star$;
it can be a new linear combination of $b_n$
oscillators.  The gauge condition
(\ref{gaugecondPsi0}) leads to an action
\begin{equation}
S_3 =  -\bra{\tilde\psi_{(4)}}
{\cal B}_{(0)} Q \ket{\psi_{(-1)}}
\equiv
 - \bra{\psi_{(3)}}  Q \ket{\psi_{(-1)}}\,, \quad  \hbox{with}  \quad
{\cal B}_{(0)}^\star  \ket{\psi_{(3)}}=0\,.
 \end{equation}
This stage of gauge fixing has given us
\begin{equation}
  {\cal B}_{(3)} = {\cal B}_{(0)}^\star\,.
\end{equation}
 Proceeding this way
 we can pick a new linear combination
 of $b_n$ oscillators for
 ${\cal B}_{(g)}$ for all $g\leq 0$ to gauge fix
 the FP ghosts $\ket{\psi_{(g)}}$. We then
 introduce FP ghosts $\ket{\psi_{(g-1)}}$
 and associated FP antighosts
 $\bra{\psi_{(3-g)}}$  which
 satisfy the condition
 ${\cal B}_{(g)}^\star \ket{\psi_{(3-g)}} = 0$.
  This shows that
\begin{equation} \label{ebbstar}
  \boxed{\phantom{\Bigl(}  {\cal B}_{(3-g)}
  = {\cal B}_{(g)}^\star \,.~}
\end{equation}
This is an important result. Since subspaces at
ghost numbers $g$ and $3-g$
are BPZ dual, the gauge-fixing condition can be
chosen freely only over ``half"
the states.

We can rewrite the gauge conditions \refb{exy1} in
a compact form
by introducing
 the gauge-fixing operator
${\cal B}$ that acts on the
full string field.  At each ghost number,
${\cal B}$ is defined to act as
the operator that imposes the relevant gauge condition.
We have
\begin{equation}
{\cal B}  = \sum_{g}   {\cal B}_{(g)}  \, \Pi_g\,,
\end{equation}
where $\Pi_g$ is the projector to the space of states of
ghost number $g$.
The gauge-fixing
condition \refb{exy1} can then
be written as  $ {\cal B} \,
\ket{\psi} = 0$
since
\begin{equation}
    {\cal B} \, \ket{\psi} =  0
    \quad \Longrightarrow\quad \sum_{g} {\cal B}_{(g)}
    \Pi_{g} \, \sum_{g'}\ket{\psi_{(g')}}    =
    \sum_{g}  {\cal B}_{(g)} \ket{\psi_{(g)}}  =0
    \quad \Longrightarrow\quad
    {\cal B}_{(g)} \, \ket{\psi_{(g)}} = 0 \text{ for all } g\,.
\end{equation}
The complete  gauge fixed free action is
given by
\begin{equation} \label{efullact}
\boxed{\phantom{\Biggl(}
S =  -{1\over 2}  \bra{\psi_{(1)}} Q
\ket{\psi_{(1)}} - \sum_{g=2}^\infty
\bra{\psi_{(g)}} Q  \ket{\psi_{(2-g)}}
=  - {1\over 2}\sum_{g=-\infty}^\infty
 \bra{\psi_{(g)}} Q \ket{\psi_{(2-g)}}
 = -{1\over 2}  \bra{\psi} Q \ket{\psi} \,,~}
\end{equation}
with the string field $\ket{\psi}$ subject to the gauge
condition $\BB\ket{\psi}=0$.\footnote{The equality
$ \bra{\psi_{(g)}} Q \ket{\psi_{(2-g)}}
=  \bra{\psi_{(2-g)}} Q \ket{\psi_{(g)}}$,
used to extend the summation range in \refb{efullact}, holds because
all string fields are Grassmann odd and $Q^\star = - Q$.}
As is well known,
the interaction term of the gauge fixed action takes the form
$-{g_o\over 3}\bra{\psi} \psi*\psi\rangle$.

\medskip
\noindent

In order to facilitate the computation of the propagator in a
general linear $b$-gauge, we shall now write down the projector
that projects onto the gauge slice.
For each ghost number $g$, we introduce
a ghost-number one operator
${\cal C}_{(g)}$ such
that\footnote{If the gauge condition
${\cal B}_{(g)}$ contains a contribution of the
form $v_0b_0$, we can choose
${\cal C}_{(g)}=v_0^{-1}c_0$. As we will see,
this is always possible for
regular
linear
$b$-gauges, as $v_0>0$ in this case.
}
\begin{equation} \label{eall}
\{ \, {\cal B}_{(g)} \,, {\cal C}_{(g)} \, \} = 1\,.
\end{equation}
This equation implies that
\begin{equation}
\{ \, {\cal B}_{(g)}^\star \,, -{\cal C}_{(g)}^\star \, \} = 1\,.
\end{equation}
Since $\BB_{(g)}^\star = \BB_{(3-g)}$, this allows us
to choose the $\CC_{(g)}$'s in such a way that
\begin{equation}
{\cal C}_{(3-g)} \equiv   - {\cal C}_{(g)}^\star \,.
\end{equation}
The
projection operator
 $\Pi_S$ into the gauge slice may now
be expressed as
\begin{equation}\label{defPiS}
\Pi_S  = \sum_g  {\cal B}_{(g)} {\cal C}_{(g)}\, \Pi_g\,.
\end{equation}
Indeed, as a consequence of \refb{eall}
this gives
\begin{equation}
\Pi_S \ket{\psi_{(g)}}  =  {\cal B}_{(g)}
{\cal C}_{(g)} \ket{\psi_{(g)}}
=\ket{\psi_{(g)}}\, ,
\ee
for a string field $\ket{\psi_{(g)}}$ satisfying the
gauge condition \refb{exy1}.
One readily verifies that $\Pi_S \Pi_S = \Pi_S$.
 To calculate the BPZ conjugate of $\Pi_S$, we first need to know the
 BPZ conjugate of the ghost number projector $\Pi_g$.
 As the inner product of a state of ghost number $g$ with a state of
 ghost number $g'$ is non-vanishing only for $g'=3-g$, we conclude that
 \begin{equation}\label{Pigstar}
    \Pi_g^\star=\Pi_{3-g}\,.
 \end{equation}
 We then have
 \begin{equation}
 \bigl({\cal B}_{(g)} {\cal C}_{(g)} \Pi_g \bigr)^\star
 = - \Pi_{g}^\star\, {\cal C}_{(g)}^\star
 {\cal B}_{(g)}^\star
 = {\cal C}_{(3-g)} {\cal B}_{(3-g)}\Pi_{3-g}
 = \bigl(1- {\cal B}_{(3-g)}
  {\cal C}_{(3-g)}\bigr) \Pi_{3-g} \, .
 \end{equation}
 Recalling the definition~(\ref{defPiS}), we obtain
\begin{equation}
\label{pistar}
\boxed{\phantom{\Bigl(} \Pi_S^\star = 1-\Pi_S\,.~}
\end{equation}
Clearly $\Pi_S^\star \Pi_S^\star = \Pi_S^\star$, so
$\Pi_S^\star$ is the orthogonal projector.

\subsection{The propagator
} \label{s2.2}
As a next step, we derive the propagator for the
class of gauge conditions discussed in \S\ref{s2.1}.
To illustrate the procedure, let us briefly review
one way of deriving the
propagator
of the free classical
string field theory.
We start out by adding a source term to the
free classical gauge-fixed
action:
\begin{equation}\label{S1source}
S_1[\psi\,, J] = -{1\over 2}  \bra{\psi_{(1)}}
 Q \ket{\psi_{(1)}}+\braket{\psi_{(1)}}{J_{(2)}}\,.
\end{equation}
 Here, the string field $\ket{\psi_{(1)}}$ is subject to the
 gauge condition $\BB_{(1)}\ket{\psi_{(1)}}=0$.
As usual, sources are arbitrary:  they are neither
killed by $Q$ nor are they
subject to gauge conditions.
We can then eliminate the classical
string field $\ket{\psi_{(1)}}$
from the action by solving its equation
of motion
\begin{equation}
     Q\ket{\psi_{(1)}}=\ket{J_{(2)}}\,.
\end{equation}
The solution to this equation for the string field
$\ket{\psi_{(1)}}$ which also obeys the gauge condition
$\BB_{(1)}\ket{\psi_{(1)}}=0$ is given by
\begin{equation}\label{classpsi1sol}
       \ket{\psi_{(1)}}  =\BQBs 11 \ket{J_{(2)}} \,,
\end{equation}
where ${\cal L}_{(g)}
 = \{ Q , {\cal B}_{(g)} \}$.
 In deriving~(\ref{classpsi1sol}) we have assumed the existence of
 the
 operators $1/{\cal L}_{(1)}$ and $1/{\cal L}_{(1)}^\star$
 which invert
 ${\cal L}_{(1)}$ and ${\cal L}_{(1)}^\star$ respectively,
   in the sense described in \S\ref{s1}.
 In \S\ref{s4} we will examine what conditions we have to
 impose on the gauge choice to be able to rigorously define
  the operators $1/{\cal L}_{(g)}$. For now we assume that such
 a suitable choice of gauge has been made.
Plugging~(\ref{classpsi1sol}) back into the
action~(\ref{S1source}) yields
\begin{equation}\label{S1sourceonly}
    S_1[\psi(J),J\,] = {1\over 2}  \BRa{J_{(2)}}
    \, \BQBs 11\,\KEt{J_{(2)}} \,.
\end{equation}
This allows us to identify the
propagator in the classical open string field theory as
\begin{equation}\label{propP2}
    {\cal P}_{(2)}= \BQBs 11 = \BQB 12\,,
\end{equation}
where we have used
the result from~(\ref{ebstar}) that
${\cal B}_{(1)}^\star={\cal B}_{(2)}$.
The subscript in ${\cal P}_{(2)}$
indicates that this propagator naturally
acts on the ghost-number two source $\ket{J_{(2)}}$.
 A propagator with the same operator structure as ${\cal P}_{(2)}$
 in~(\ref{propP2}) first appeared in~\cite{preit}.
For the case of Schnabl gauge, the above propagator was
first mentioned in~\cite{0511286} and it was
used to calculate the off-shell Veneziano amplitude
in~\cite{0609047,0708.2591}.

It is now easy to generalize this
construction to the complete
gauge-fixed free action \refb{efullact}.
We include
sources $\ket{J_{(3-g)}}$
for gauge-fixed
string fields $\ket{\psi_{(g)}}$ of all
ghost numbers
and obtain
\begin{equation}\label{Sgsource}
    S[\psi,\,J\,] = -{1\over 2} \sum_{g=-\infty}^{\infty}
    \bra{\psi_{(g)}} Q \ket{\psi_{(2-g)}}
    +\sum_{g=-\infty}^{\infty}\braket{\psi_{(g)}}{{J_{(3-g)}}} \,.
\end{equation}
The equation of motion for $\ket{\psi_{(g)}}$ now reads
\begin{equation}
     Q\ket{\psi_{(2-g)}}=\ket{{J_{(3-g)}}}\,.
\end{equation}
It is again straightforward to determine
the
string field $\ket{\psi_{(2-g)}}$ which solves this equation and
also satisfies
the gauge condition $\BB_{(2-g)}\ket{\psi_{(2-g)}}=0$.
We obtain
\begin{equation}\label{classpsigsol}
       \ket{\psi_{(2-g)}}   =
           \BQBs {2-g}g \ket{J_{(3-g)}} \,,
\end{equation}
or, equivalently,
\begin{equation}
          \bra{\psi_{(g)}}
          =\bra{J_{(1+g)}}\BQBs {2-g}g  \,.
\end{equation}
Plugging these results back  into
(\ref{Sgsource}) yields
\begin{equation}\label{Sgsourceonly}
    S[\psi(J),J\,] = \frac{1}{2}\sum_{g=-\infty}^{\infty}
    \BRa{J_{(1+g)}}\BQBs {2-g}g \KEt{{J_{(3-g)}}}
    = \frac{1}{2}\sum_{g=-\infty}^{\infty}
    \BRa{J_{(4-g)}}\BQB {g-1}g \KEt{{J_{(g)}}} \,,
\end{equation}
where we used \refb{ebbstar} in obtaining the second equality.
We can now
identify the propagator ${\cal P}_{(g)}$
acting on the source $\ket{J_{(g)}}$ of ghost number $g$ as
\begin{equation}\label{propagatorg}
    \boxed{ \phantom{\Biggl(} {\cal P}_{(g)}   =
       \BQB {g-1}g \,.\,}
\end{equation}
Alternative expressions
obtained by using the BPZ conjugation property~(\ref{ebbstar}) are
\begin{equation}\label{propagatorg99}
    {\cal P}_{(g)} =\,\, \BQBs {g-1}{3-g}\,\,=\,\,\BsQB {4-g}{g}
   \, \, =\,\,\BsQBs{4-g}{3-g}\,.
\end{equation}

We  can
simplify notation by combining all sources
$\ket{J_{(g)}}$
into a single source
 \begin{equation}
    \ket{J}\equiv \sum_{g=-\infty}^{\infty}\ket{J_{(g)}} \,,
\end{equation}
just as we did for the
 gauge-fixed
string field $\ket{\psi}$ in~(\ref{edeftot}).
Let us furthermore define the full propagator
${\cal P}$ as the operator whose action on a
subspace of ghost number $g$ is given by ${\cal P}_{(g)}$, i.e.
\begin{equation}\label{propagator}
 \boxed{ \phantom{\Biggl(}   {\cal P}\equiv\sum_{g=-\infty}^{\infty}
    {\cal P}_{(g)} \, \Pi_{g} \,.~}
\end{equation}
Then the elimination of
$\ket{\psi}$
from the free action can be conveniently summarized as
\begin{equation}\label{Sgshort}
 S[\psi,J\,] = -{1\over 2} \bra{\psi} Q\ket{\psi}
    +\braket{\psi}{J} \quad \to
    \quad S[\psi(J), J\,]
    =\frac{1}{2}\,\bra{J}{\cal P}\ket{J} \,.
\end{equation}

Equations
\refb{propagatorg} and \refb{propagator}
give the full propagator ${\cal P}$ for general linear $b$-gauges.
The propagator acts
differently on states of different ghost number.
This is not surprising,
considering that for generic linear $b$-gauges it is impossible to
impose the same gauge condition on
states of all ghost numbers.

\subsection{Properties of the propagator} \label{s2.3}

Let us now turn to study the algebraic
properties of the full propagator.
We claim that ${\cal P}$ satisfies the important relation
\begin{equation}\label{QP=1}
    \boxed{ \phantom{\biggl(} \{Q,{\cal P}\}=1 \,. \,}
\end{equation}
We can prove this property as follows:
\begin{eqnarray}\label{QP=1proof}
    \{Q,{\cal P}\} \, \Pi_g
    &=& \left( Q \PP_{(g)}+\PP_{(g+1)} Q\right) \Pi_g
    =\Biggl(Q \,
    \BQB {g-1}g + \BQB {g}{g+1} \,Q\Biggr)\Pi_g \nonumber \\
    &=&
      \Biggl(Q \,
    \frac{{\cal B}_{(g)}}{{\cal L}_{(g)}}
    +\frac{{\cal B}_{(g)}}{{\cal L}_{(g)}}
    \, Q\Biggr)\Pi_g  = \Pi_g \,.
\end{eqnarray}
 Here we have again assumed that the
 operator $1/{\cal L}_{(g)}$ can be
 defined rigorously
  in the sense described in \S\ref{s1}
 for the linear $b$-gauge under consideration.
Equation (\ref{QP=1proof})
shows that
$\{Q,{\cal P}\}=1$
holds on all subspaces of fixed
ghost number $g$, and it thus holds in general.
Notice that ${\cal P}_{(g)}$,
regarded as an operator acting on
states of arbitrary ghost number,
generically does not satisfy the same
property:
\begin{equation}
    \{Q,{\cal P}_{(g)}\}=Q \, \frac{{\cal B}_{(g)}}
    {{\cal L}_{(g)}}+\frac{{\cal B}_{(g-1)}}{{\cal L}_{(g-1)}} \, Q
    \neq 1 \qquad \text{ if }\quad
    {\cal B}_{(g)}\neq {\cal B}_{(g-1)} \,.
\end{equation}
It is precisely the property~(\ref{QP=1})
which will allow us to prove the
decoupling of pure-gauge states and the
correctness of on-shell amplitudes in \S\ref{s3}.

The propagator ${\cal P}$ is BPZ-invariant,
\begin{equation}\label{P*=P}
   \boxed{ \phantom{\Bigl(}{\cal P}^\star={\cal P}\,. ~ }
\end{equation}
 Indeed, using~(\ref{ebbstar}) and~(\ref{Pigstar}) we obtain
 \begin{equation}\label{}
    \bigl(\PP_{(g)}\,\Pi_g\bigr)^\star
    =\Pi_{g}^\star\,\BsQBs{g}{g-1}
    = \Pi_{3-g} \, \BQB{3-g}{4-g}
    =\BQB{3-g}{4-g}\,\Pi_{4-g}\,.
 \end{equation}
 Recalling the definition~(\ref{propagator}) of the propagator, this establishes $\PP^\star=\PP$.

In addition, the propagator  satisfies a set of simple
properties related to the projection operator $\Pi_S$ to
the gauge slice:
\begin{equation}
   \Pi_S\, {\cal P}=\PP\, \Pi_S^\star= {\cal P}, \qquad
   \Pi_S^\star\, \PP = \PP\, \Pi_S =0\,.
\end{equation}
These equations are readily checked acting on subspaces
of fixed ghost number,  using  the
definitions of ${\cal P}_{(g)}$
and $\Pi_S$, and eq.\refb{pistar}.

 It is convenient to introduce the
 gauge-fixed kinetic operator $\cal{K}$, given by
 \begin{equation}\label{kinope}
    {\cal K} \equiv \Pi_S^\star \, Q \, \Pi_S \,,\qquad
    {\cal K}^\star = - {\cal K} \,.
 \end{equation}
Using this
and  $\{Q,{\cal P}\}=1$
we then find
\begin{equation}
\begin{split}
   {\cal P} \, {\cal K} ={\cal P} \, \Pi_S^\star \,  Q \,  \Pi_S
   ={\cal P} \, Q \, \Pi_S
   =(1-Q \, {\cal P}) \, \Pi_S
   = \Pi_S \,.
\end{split}
\end{equation}
This and the BPZ conjugate relation are
\begin{equation}
   \boxed{ \phantom{\Bigl(} {\cal P} \, {\cal K}=\Pi_S \,,\quad
     {\cal K} \, {\cal P}=\Pi_S^\star \,. ~ }
\end{equation}
This shows that, as expected,
the propagator inverts the
gauge-fixed kinetic operator on the gauge slice.

\subsection{Constraints on linear
$b$-gauges} \label{sreality}

So far in our analysis
we have not imposed any restriction on the linear combinations
of $b_n$ oscillators which
define the operators $\BB_{(g)}$.
The vector field $v(\xi)$ associated
with any of the operators $\BB_{(g)}$ through the relations
\begin{equation}\label{vofBg}
\BB_{(g)} =  \int {d\xi\over 2\pi i} \,
v(\xi) b(\xi)
= \sum_n v_n b_n \,, \qquad
v(\xi) = \sum_n v_n \xi^{n+1}\,,
\end{equation}
was taken to be completely arbitrary.
In this subsection we will examine what
constraints we need to impose
on the coefficients $v_n$ to obtain a
physically reasonable gauge choice.

First of all, in order to facilitate the analysis
of string perturbation theory we require
that  the string field theory Feynman diagrams
represent correlation functions on Riemann surfaces.
For this we require
the validity of the Schwinger representation of the
factors of $1/\LL_{(g)}$
in the propagator:
\be \label{erepr}
{1\over \LL_{(g)}} =\lim_{\Lambda_{(g)}\to\infty}
\int_0^{\Lambda_{(g)}} ds\, e^{-s \LL_{(g)}}\, .
\ee
Furthermore, the insertion of $e^{-s\LL_{(g)}}$ into a
correlation function
must represent the insertion of a piece of world
sheet to the Riemann surface that
represents the rest of the diagram. For this $\LL_{(g)}$
must generate a conformal transformation.
In open string theory a conformal transformation
$\delta \xi\propto\, v(\xi)$ is generated~by
\be \label{egene1}
\int_C \left({d\xi\over 2\pi i} \, v(\xi)\, T(\xi)
+ {d\bar \xi\over 2\pi i}
 \, \,\overline{v(\xi)}\,\,
\overline{T(\xi)} \right)\, ,
\ee
where $C$ denotes
 the unit semicircle in the upper-half
plane and
bars indicate complex conjugation.
Replacing $v(\xi)$ by $\xi^{n+1}$ we get
the generators of conformal
transformation:
\be \label{egene}
L_n = \int_C \left({d\xi\over 2\pi i}  \xi^{n+1} \, T(\xi)
+ {d\bar \xi\over 2\pi i} \bar\xi^{n+1}\,
\overline{T(\xi)} \right)\, .
\ee
This gives
\be \label{egene2}
\LL_{(g)}=\sum_n v_n L_n =
\int_C \left({d\xi\over 2\pi i} v(\xi) T(\xi)
+ {d\bar \xi\over 2\pi i} v(\bar\xi)\,
\overline{T(\xi)} \right)\, .
\ee
This does not
have the form of the generator \refb{egene1}
unless
\begin{equation}\label{condreal1}
     \overline{v(\xi)}=v(\bar\xi)\,.
\end{equation}
Thus, in order
that the Feynman
diagrams generated by open string field theory have a
direct Riemann surface interpretation
we must require that
the coefficients $v_n$  be
real,\footnote{We can try to define the results for complex
$v_n$ by
analytic continuation of the real $v_n$ results.
This trick was used in \cite{preit}
to discuss the gauge condition $(b_1+b_{-1})\ket{\psi}=0$.
We shall not consider
this possibility here.}
\i.e.\ $v(\xi)$ to be real on the real axis.

Even when \refb{condreal1} holds and the insertion of
$e^{-s\LL_{(g)}}$ has a Riemann surface interpretation,
eq.\,\refb{erepr} may fail to provide the correct definition
of $1/\LL_{(g)}$ due to a non-vanishing contribution from
the upper limit of integration.
This requirement will be
analyzed in detail in section~\S\ref{s4}.
It leads to
 condition~(\ref{ealt2old}) which requires the vector field $v(\xi)$
 to be analytic in some neighborhood of the unit circle $|\xi|=1$ and
 to satisfy
 \be \label{ealt2old2}
 v_\perp (\xi) \equiv \Re\bigl(\bar\xi v(\xi)\bigr)
 > 0 \quad
 \hbox{for $|\xi|=1$} \, .
 \ee

Secondly, in order that
the open string field theory action is real, describing a
unitary quantum theory,
the string field
and the interaction vertices
must satisfy certain reality
conditions.
The reality condition on the string
field is easily stated~\cite{Gaberdiel:1997ia}:
the combined operations of BPZ conjugation
and hermitian conjugation (HC) -- called star conjugation --
must leave the string field invariant.
In open string field theory the interaction term must also be real.
If the string field is real,
the interaction term
is real once the coordinate
systems around the punctures
on the Riemann surface associated with the interaction
vertex
satisfy a reality condition.
The interaction vertex of Witten's open string
field theory  satisfies this condition.

Therefore, when we impose linear $b$-gauge
conditions we must make
sure that this can be done consistently with the constraint of
real string fields.  Since the total effect
of BPZ followed by HC does not
change the ghost number, we can analyze
the condition on string fields of
fixed ghost number.
Consider the gauge condition
$\BB_{(g)}\ket{\psi_{(g)}} =0$,
with $\BB_{(g)}$ related to a vector field $v(\xi)$
through the relation~(\ref{vofBg}).
In order to
impose the reality condition we need
that if $\ket{\psi_{(g)}}$ satisfies
the gauge condition then
the star-conjugate of $\ket{\psi_{(g)}}$
automatically satisfies
the gauge
condition --
this allows us to
form the  linear combination required for reality.
For this we must have
$( \BB_{(g)}^\star)^\dagger \propto
\BB_{(g)}$ with $\dagger$ denoting hermitian conjugation.
Since the operation of
star conjugation is an involution one can only have
\be \label{ebeqn}
( \BB_{(g)}^\star)^\dagger  = e^{i\alpha}\,
 \BB_{(g)}\, ,
 \ee
with $\alpha$ real.
Recalling that $(b_n)^\star = (-1)^n b_{-n}$ and
$(b_n)^\dagger = b_{-n}$, a short calculation shows that
\begin{equation}\label{defvstar}
\begin{split}
\BB_{(g)}^\star &\equiv\int {d\xi\over 2\pi i}
 \, b(\xi) v^\star(\xi) \,, \quad
\hbox{with} \quad  v^\star (\xi) =
- \xi^2 v \bigl( -{1/ \xi} \bigr) \\[0.5ex]
\BB_{(g)}^\dagger &\equiv\int
{d\xi\over 2\pi i} \, b(\xi) v^\dagger(\xi) \,, \quad
\hbox{with} \quad  v^\dagger (\xi) = \xi^2\,
\overline{v \bigl( {1/ \bar \xi} \,\bigr)}\,.
\end{split}
\end{equation}
Thus \refb{ebeqn} holds
if $(v^\star)^\dagger =
e^{i\alpha} v$.
Since
\begin{equation}
\label{star_of_vector}
( v^\star (\xi) )^\dagger = \bigl( -\xi^2 v(-1/\xi) \bigr)^\dagger
= \xi^2 \overline{ \bigl( - {\bar \xi}^{-2}  \, v ( - \bar \xi) \bigr)}
= - \,\,\overline{ v(-\bar\xi)} \,,
\end{equation}
we can rewrite the condition on $v(\xi)$ as
\begin{equation}
\label{gauge_reality_consistency}
\,e^{i\alpha} v(\xi) = -\,\, \overline{ \phantom{a^{a^2}}\hskip-12pt
v(-\bar\xi)} \,, \quad \hbox{for some real $\alpha$}\, .
\end{equation}

Recalling that we required $v(\xi)$ to be
real on the real axis, only
$e^{i\alpha}=\mp1$
are allowed
in~(\ref{gauge_reality_consistency}).
Combining this with the condition
\refb{condreal1} we
conclude that
the vector field $v$ has to be either
 even or odd under $\xi\to-\xi$:
\begin{equation}\label{}
    v(-\xi) =
     \pm \,
    v(\xi) \,.
\end{equation}
 It is
easy to see that
the choice $v(-\xi)=v(\xi)$ is not compatible with
conditions~(\ref{condreal1}) and~(\ref{ealt2old2}).
To prove this  assume that
condition~(\ref{ealt2old2}) is
satisfied for some $\xi$ on the upper-half unit circle:
\begin{equation}\label{}
   \Re\bigl(\bar\xi\,v(\xi)\bigr)>0\,.
\end{equation}
Using~(\ref{condreal1}) and $v(-\xi)=v(\xi)$,
it immediately follows that
\begin{equation}\label{}
     \Re\bigl(\overline{(-\bar\xi)} v(-\bar\xi)\bigr)
     =\Re\bigl((-\bar\xi)\overline{v(-\bar\xi)}\bigr)
     =\Re\bigl((-\bar\xi) v(-\xi)\bigr)
     =-\Re\bigl(\bar\xi v(\xi)\bigr)<0\,,
\end{equation}
in contradiction with
condition~(\ref{ealt2old2})
for $-\bar\xi$.
Thus we conclude that
 physically reasonable gauges must satisfy
\begin{equation}\label{049irjfgi84}
     \overline{v(\xi)}=v(\bar \xi)\,, \quad v(-\xi) =-v(\xi)\,,
\end{equation}
and thus
\begin{equation} \label{eonepossible}
  \boxed{\phantom{\Biggl(}
  v(\xi) = \sum_{k\in\mathbb{Z}} v_{2k}\,  \xi^{2k+1}  \quad
  \hbox{with} ~v_{2k}\in \mathbb{R} \, .~}
\end{equation}

It should be noted
that the conditions derived so far are consistent with
\refb{ebbstar} --
if $v(\xi)$ satisfies \refb{ealt2old2} and
the additional conditions \refb{049irjfgi84},
so does the dual vector $v^\star(\xi)$.
Indeed, on the unit circle
\be
\bar\xi v^\star (\xi) = -\xi v (-1/\xi)  = \xi v(1/\xi) = \xi v(\bar \xi) = \overline
{\bar \xi \, v(\xi)}\,.
\ee
It follows that $\Re\bigl( \bar\xi v^\star (\xi)\bigr) = \Re \bigl({\bar \xi \, v(\xi)}
\bigr) > 0$, as we wanted to show.  It is straightforward to
show that $v^\star(\xi)$ satisfies~\refb{049irjfgi84}.

One can examine the constraint~(\ref{ealt2old2})
more explicitly using the Laurent expansion of the
vector $v(\xi)$.
Writing $\xi=e^{i\theta}$ we find
\begin{equation}\label{cos2k}
    v_\perp(e^{i\theta})=v_0+\sum_{k\neq0}
    v_{2k}\cos(2k\theta)>0\,.
\end{equation}
Thus
the average of $v_\perp(e^{i\theta})$ over
$0\leq\theta\leq\pi$ is given by
\begin{equation}\label{}
    \frac{1}{\pi}\int_0^\pi d\theta\,v_\perp(e^{i\theta})=v_0\,,
\end{equation}
leading to the constraint
\begin{equation}\label{}
    \boxed{\phantom{\biggl(}~v_0>0\,.~ }
\end{equation}
All operators $\BB_{(g)}$ must contain a component along $b_0$
with positive coefficient.
It also follows from~(\ref{cos2k}) that
\begin{equation}\label{4598gn}
v_0>\sum_{k\neq0}|v_{2k}|
\end{equation}
is sufficient (but not necessary!)
for
condition~(\ref{ealt2old2}) to be satisfied.
It is useful to check that \refb{4598gn} is
 not satisfied
for Schnabl gauge. In this gauge (\ref{new15}) tells us that the only
nonvanishing coefficients are
\be
v_0=1\,, \quad  v_{2k} = {2(-1)^{k+1}\over 4 k^2 -1}\,,\quad  k = 1,2, \ldots
\ee
A short calculation gives
\be
\label{kchthaitw}
\sum_{k=1}^\infty  |v_{2k} |
= 2 \sum_{k=1}^\infty {1\over 4k^2 -1} = 1= v_0\,,
\ee
showing that  \refb{4598gn} is
marginally violated.
This failure is in fact
related to the vanishing of $v(\xi)$ for $\xi=i$:
\be
v(\xi) =  i \Bigl(v_0 + \sum_{k=1}^\infty v_{2k} (-1)^{k} \Bigr)
=  i \Bigl(v_0 - \sum_{k=1}^\infty |v_{2k} | \Bigr)  = 0\,.
\ee
The vanishing of the vector at any point on the circle means that
the conditions for a regular gauge are not satisfied.

\subsection{Examples} \label{s2.4}
To define a specific linear $b$-gauge,
we need to choose
a linear combination of oscillators $b_n$
for each ${\cal B}_{(g)}$ with $g\leq 1$.
The remaining ${\cal B}_{(g)}$ are then fully
determined through the
relation~(\ref{ebbstar}),
${\cal B}_{(3-g)}={\cal B}^\star_{(g)}$.
The simplest linear $b$-gauge is Siegel gauge:
${\cal B}_{(g)}=b_0$.
As the Siegel gauge condition is BPZ
invariant, we can impose the same condition
on string fields of all ghost numbers.
Schnabl gauge corresponds to the choice
${\cal B}_{(1)}=B$ for classical
string fields,
with $B$ defined in~(\ref{new15}).
Geometrically, $B$ can be understood
as the zero mode of the antighost in the sliver frame:
\be \label{ebdef}
B = f^{-1}\circ \oint {dz\over 2\pi i} z b(z)
=  \oint {d\xi\over 2\pi i}  {f(\xi)\over f'(\xi)} b(\xi)\,
=  \oint {d\xi\over 2\pi i}  v(\xi) b(\xi) \,,\ee
where $\circ$ denotes a conformal transformation,
the sliver frame coordinate $z=f(\xi)$ is
given by
\begin{equation} \label{efdef}
     f(\xi)=\frac{2}{\pi}\tan^{-1} \xi\, ,
     \end{equation}
and
\begin{equation}\label{slivv}
     v(\xi) = \frac{f(\xi)}{f'(\xi)} = (1+ \xi^2) \tan^{-1} \xi
\end{equation}
is the vector field associated with $B$.
The function $f$ maps the point $\xi=i$ to infinity.
This property implies that the sliver, regarded as a
surface state with local coordinates
 on the upper half plane
defined through the map $f$, is a projector. Conversely, any map
$f(\xi)$ that sends the point $i$ to infinity can be used
to describe a projector. A gauge choice is called
a \emph{projector gauge} if
$\BB_{(1)}$ is defined
 just like the operator $B$ in \refb{ebdef},
 but with $f(\xi)$ describing an arbitrary projector.
Thus Schnabl gauge is a projector gauge.
For any projector gauge
$f(\xi)$ diverges at $\xi=i$
and so does $\ln f(\xi)$ and its derivative $f'(\xi)/f(\xi)$.
It follows that the associated vector field
$v(\xi)=f(\xi)/f'(\xi)$ vanishes
at $\xi=i$ and hence fails to satisfy condition
\refb{ealt2old}.
 Thus projector gauges are not
regular
gauges in the sense described in \S\ref{s1}.

There is a natural
one-parameter family of
 regular
gauges ${\cal B}_{(1)}=B^\lambda$
parameterized by
$0< \lambda<\infty$
which interpolates
between Siegel and Schnabl gauge.
$B^\lambda$ is defined by
\begin{equation}
    B^\lambda \equiv e^{\lambda L_0} \, B \, e^{-\lambda L_0} \,,
    \qquad\text{ with }0< \lambda<\infty \,.
\end{equation}
Since $B$ is
the sum of $b_0$ and
a linear combination of $b_n$'s with
$n>0$, the relation
\begin{equation}
\label{mode-bn_change}
      e^{\lambda L_0} \, b_n \, e^{-\lambda L_0}= e^{-\lambda n} b_n \,
\end{equation}
ensures that we recover Siegel gauge in the limit
$\lambda\to\infty$:
\begin{equation}
    \lim_{\lambda\to\infty}B^\lambda = b_0 \,,
\end{equation}
Schnabl gauge on the other hand
 is not a regular linear $b$-gauge and
corresponds to $\lambda \to 0$:
\begin{equation}
     B = \lim_{\lambda\to0} B^{\lambda} \,.
\end{equation}

The operator $B^\lambda$ is also the zero mode of the antighost
field  in a certain conformal frame determined up to a real scaling.
To determine such a frame
$z= f^\lambda (\xi)$ we note that the associated vector field
$v^\lambda (\xi)$ differs
in a simple manner
from the sliver vector field
 $v(\xi)$ of~(\ref{slivv}).
If we
expand
\begin{equation}
v(\xi) = \sum_{k\in\mathbb{Z}}  v_{2k} \, \xi^{2k+1}\,,
\end{equation}
then equation \refb{mode-bn_change}  tells us that
\begin{equation}
\label{vectorvlambda}
v^\lambda (\xi)  =
 \sum_{k\in\mathbb{Z}}  v_{2k} \, e^{-2 k  \lambda }  \xi^{2k+1}
 = e^\lambda \sum_{k\in\mathbb{Z}} v_{2k} \,
 (e^{-\lambda}  \xi)^{2k+1}
= e^\lambda\, v( e^{-\lambda} \xi) \,.
\end{equation}
It is now simple to verify that
\begin{equation}
\label{functionlambda}
f^\lambda (\xi) =  f (e^{-\lambda} \xi) =  {2\over \pi} \,
\tan^{-1} (e^{-\lambda} \xi) \,,
\end{equation}
satisfies the expected relation
$f^\lambda (\xi)/ {f^\lambda}' (\xi) = v^\lambda
(\xi)$.
For any $\lambda >0$ the  coordinate curve $f^\lambda (e^{i\theta})$,
$\theta \in [0, \pi]$ is smooth and reaches a maximum finite
height for $\theta = \pi/2$, as shown in  Fig.~\ref{yyy}.\footnote{
 In (\ref{functionlambda}) we chose the normalization $2/\pi$
 to reproduce the sliver frame coordinate~(\ref{efdef})
 in the limit $\lambda\to0$.
 Alternatively, the normalization $1/\tan^{-1}(e^{-\lambda})$ is
 convenient to study the Siegel limit
 $\lambda\to\infty$, because we have
 $f^\lambda (\xi)=\tan^{-1} (e^{-\lambda} \xi)/
 \tan^{-1}(e^{-\lambda})=\xi+\ord{e^{-\lambda}}$ in this case.}
\begin{figure}
\centerline{\epsfig{figure=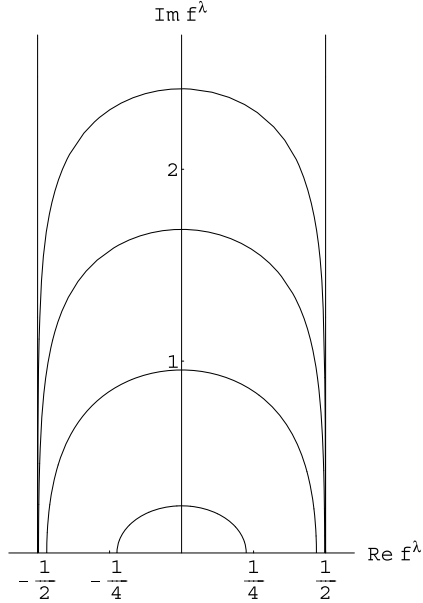, height=8cm}}
\caption{The coordinate curve
$f^\lambda (e^{i\theta})$,
$\theta \in [0, \pi]$
associated with $B^\lambda$,
plotted for $\lambda=1,\,0.1,\,0.01,\,0.001$, and $\lambda=0$.
The latter is the sliver frame and the coordinate
curve consists of straight vertical lines
that reach $i\infty$ for
$\theta=\pi/2$.
}
\label{yyy}
\end{figure}

The vector field $v^\lambda(\xi)$ given in
\refb{vectorvlambda}
satisfies the conditions \refb{ealt2}.
The only
nontrivial condition is the
one rephrased in
\refb{cos2k}.
From~(\ref{vectorvlambda})
we see that the expansion coefficients for the vector fields
$v^\lambda$ and $v$ are related by
\be
v_{2k}^\lambda  = e^{-2k\lambda}  v_{2k} \,,  \quad k = 0 ,1, 2, \ldots \,.
\ee
It then follows that for any $\lambda >0$
\be
\sum_{k=1}^\infty |v_{2k}^\lambda|  <  \sum_{k=1}^\infty |v_{2k}| = 1\,,
\ee
after use of \refb{kchthaitw}.  This shows that the vector $v^\lambda$ satisfies \refb{4598gn}, which suffices
for a regular
gauge.
The
$B^\lambda$ gauge is not a projector
gauge for $\lambda>0$ and
does not exhibit the problematic
properties of Schnabl gauge.
The $\lambda$ parameter  allows
us to
interpolate between Schnabl and Siegel gauge.
It may also allow
us to regularize
and {\em define}
amplitudes in Schnabl gauge as the limit
$\lambda\to 0$ of amplitudes in the $B^\lambda$ gauge.

\medskip
Neither Schnabl gauge nor the $B^\lambda$
gauges impose a BPZ invariant gauge
condition on the classical
string field
$\ket{\psi_{(1)}}$. There is  therefore no
preferred choice of gauge conditions on the
ghost sector string fields.
Let us discuss one possible
assignment of gauge conditions which we will call
\emph{alternating gauge}. In alternating gauge,
we apply
the classical
gauge condition ${\cal B}_{(1)}
\ket{\psi_{(1)}}=0$
to all  string fields of odd ghost number $g$,
\begin{equation}
     {\cal B}_{(1)}\ket{\psi_{(g)}}=0   \qquad\text{for $g$ odd} \,.
\end{equation}
This exhausts our freedom to choose
conditions.
The relation
${\cal B}_{(3-g)}={\cal B}^\star_{(g)}$
forces us to
assign the BPZ conjugate condition on
string fields of even ghost number $g$:
\begin{equation}
     {\cal B}_{(1)}^\star\ket{\psi_{(g)}}=0
     \qquad\text{for $g$ even} \,.
\end{equation}
 Let us denote the  projectors onto states of even and
 odd ghost numbers by $\Pi_+$ and $\Pi_-$, respectively.
Then we can state the gauge condition as
\begin{equation}
     {\cal B}\ket{\psi}=0\,, \quad\text{ with }\quad
     {\cal B} \equiv {\cal B}_{(1)}\, \Pi_- +
     {\cal B}_{(1)}^\star \, \Pi_+ \,.
\end{equation}
The propagator in alternating gauge is readily seen to be
given by
\begin{equation}
\begin{split}
    {\cal P} &= \frac{{\cal B}_{(1)}}{{{\cal L}_{(1)}}}
    \, Q \, \frac{{\cal B}_{(1)}^\star}{{\cal L}_{(1)}^\star}
    \,\Pi_+
    + \frac{{\cal B}_{(1)}^\star}{{\cal L}_{(1)}^\star}
    \, Q \, \frac{{\cal B}_{(1)}}{{\cal L}_{(1)}}\,\Pi_- \,.
\end{split}
\end{equation}

\section{Analysis of  on-shell amplitudes
} \label{s3}
\setcounter{equation}{0}

In this section we shall analyze the on-shell amplitudes in a
general linear $b$-gauge. In \S\ref{spure} we give a simple
proof of the decoupling of pure-gauge states. In
\S\ref{ssig} we prove the equality of on-shell amplitudes in
a general linear $b$-gauge and the Siegel gauge. Since the
latter is known to reproduce correctly the Polyakov amplitudes
in open string theory, this establishes that the on-shell amplitudes
in a linear $b$-gauge give the correct S-matrix of open string
theory. The proofs in this section rely on the validity of the relation
\begin{equation}\label{assump}
\left\{ Q,  \PP\right\}= 1\, .
\end{equation}
In \S\ref{s4} and \S\ref{s5}
we will carefully analyze this relation
by regularizing the operators
 $1/{\LL_{(g)}}$ that enter in the definition of $\PP$.
  We will then determine
the conditions
that  we need to impose for all correction terms
 to be localized at the boundary of open string moduli space in the limit
when we remove the regularization.
We will find
that~(\ref{assump})
can be made rigorous for gauge
choices which are regular gauges as defined in \S\ref{s1}.

\subsection{Decoupling of pure gauge states} \label{spure}

\begin{figure}[p]
\centerline{\epsfig{figure=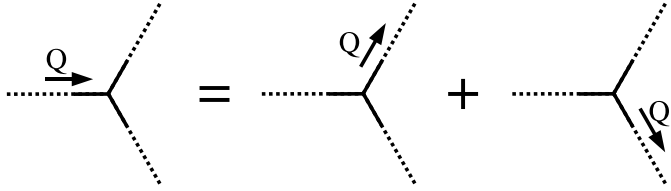, height=2.5cm}}
\caption{Diagram illustrating
the movement of $Q$ through a vertex.}
\label{fid1}
\end{figure}
\begin{figure}[p]
\centerline{\epsfig{figure=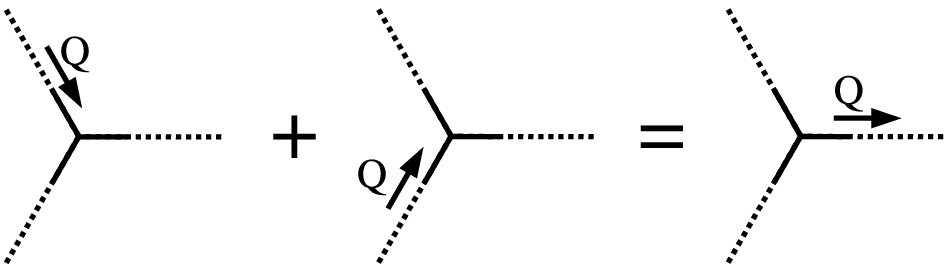, height=2.5cm}}
\caption{Another
diagram illustrating the movement of $Q$ through a vertex.}
\label{fid1.5}
\end{figure}
\begin{figure}[p]
\centerline{\epsfig{figure=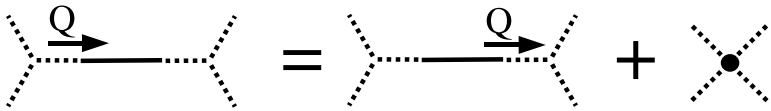, height=1.5cm}}
\caption{Diagram illustrating the movement of $Q$ through a
propagator.} \label{fid2}
\end{figure}
 \begin{figure}[p]
 \centerline{\epsfig{figure=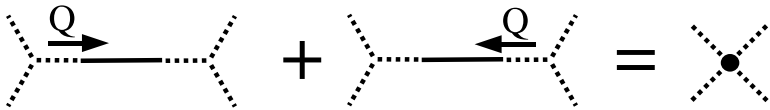, height=1.5cm}}
 \caption{Diagram illustrating $Q$ collapsing a
 propagator.} \label{fid2b}
 \end{figure}
\begin{figure}[p]
\centerline{\epsfig{figure=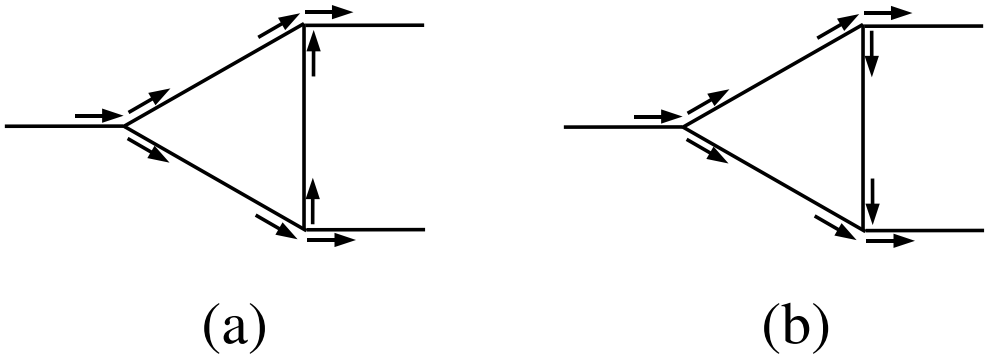, height=5cm}}
\caption{Moving $Q$ through a
diagram. There are many different ways of
moving $Q$ through a diagram, as shown in figures (a)
and (b); but the final result is independent of this
choice.} \label{fmove}
\end{figure}

Consider an on-shell amplitude where every external state is
BRST closed and, furthermore,
 one of the external states
is pure gauge, \i.e.\ has the form $Q\ket{\chi}$ for some
ghost-number zero state $\ket\chi$. In this case we can move the $Q$
through the various propagators and vertices of a
Feynman  diagram contributing to this
amplitude using the relations
\begin{eqnarray}
Q^{(1)} \, \ket{V_{123}}
&=& {}-\left(Q^{(2)} + Q^{(3)}\right) \ket{V_{123}}\, , \label{ex3}\\
\left( Q^{(1)} + Q^{(2)} \right)\, \ket{V_{123}}
&=& {}- Q^{(3)}  \ket{V_{123}}\, , \\
Q \, \PP &=& - \PP \, Q + 1 \label{ex4}\\
Q \, \PP+ \PP \, Q &=&  1 \label{ex4b}
\, .
\end{eqnarray}
Here $\ket{V_{123}}$ denotes the three string vertex.
The diagrammatic
representations of these three identities are shown in
Figs.~\ref{fid1}, \ref{fid1.5},
\ref{fid2}, and \ref{fid2b}.
An example of how $Q$ moves
through a given diagram has been shown in Fig.~\ref{fmove}.
In fact there are many different orders in which we can move
$Q$ through a given diagram, as shown in
Figs.\ref{fmove}(a) and \ref{fmove}(b),  but the final result
is independent of this choice and so for each diagram we
make a fixed choice.
Eqs.\refb{ex4}
 and \refb{ex4b} show
that during the process of moving $Q$ through a propagator
we are left with an extra contribution where the propagator is
replaced by unity. We label it by a collapsed propagator, \i.e\ a
four point vertex.
It is clear from the identities~(\ref{ex3})-(\ref{ex4b})
and their diagrammatic representations
Fig.\ref{fid1}-Fig.\ref{fid2b}
that the BRST operator moves through the
diagram until it hits an external state,
or until it collapses an internal propagator. The contributions
from hitting external states
 vanish, because the external states
 are BRST closed.
 Therefore we are left with the contributions from
 collapsing propagators.
 A diagram with $n$ internal propagators gives rise to
 $n$ such terms. In each term,
 one internal line of the diagram has collapsed,
while all other lines have the original propagator $\PP$.
We now combine
contributions from different Feynman diagrams.
In this case each diagram with a collapsed
propagator arises in two different ways, one where the
collapsed
propagator appears as a t-channel propagator, and another
where it
appears
as an s-channel propagator
(see Fig.~\ref{f1}).\footnote{Since
in the Riemann surface interpretation of string field theory
Feynman diagrams each line is blown up to a strip, the cyclic
ordering of the labels $i$, $j$, $k$, $l$ is important. Thus for
example if we exchange the labels $i$ and $j$ in the left-most
diagram of Fig.~\ref{f1} then it would be regarded as a different
string Feynman diagram. For this reason
a diagram
with a collapsed $u$-channel propagator has a different structure and needs to be combined with another diagram carrying a different cyclic
ordering of the labels $i$, $j$, $k$, $l$ from the one shown
in Fig.~\ref{f1}.}
When this propagator collapses, the t-channel
and s-channel diagrams are
indistinguishable, and their contributions cancel
because they come with opposite
signs -- a result familiar
from the proof of decoupling
of pure gauge states in ordinary Siegel gauge amplitudes.
Thus the
diagrams with collapsed propagators cancel pairwise.
This finishes our proof of decoupling of pure gauge states
in on-shell amplitudes.

\begin{figure}
\centerline{
\epsfig{figure=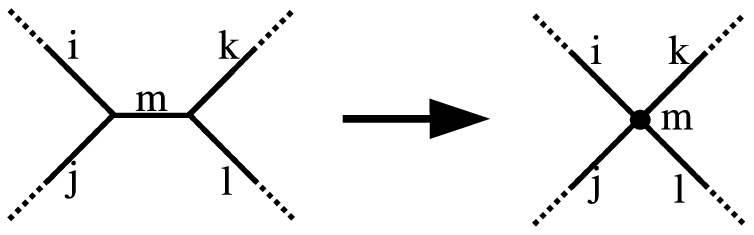, height=2.5cm}
\hskip 2cm
\epsfig{figure=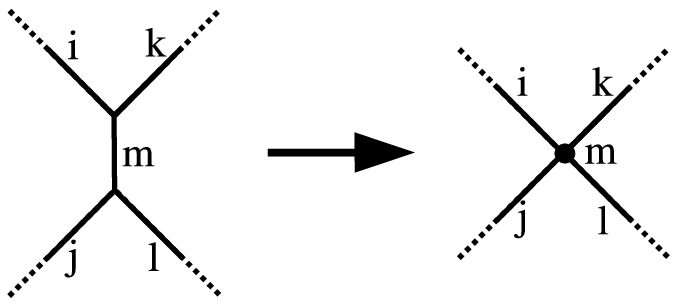, height=3cm}}
\caption{The cancelation between $s$- and $t$-channel diagrams
with collapsed propagators.} \label{f1}
\end{figure}

While the proof itself was straight-forward, it is important to
identify the main ingredient of the proof.
It is in fact eq.\refb{ex4} that tells us that when $Q$ passes
through a propagator it leaves behind a contribution that is
unity. Had this been a non-trivial operator in the CFT, the
cancellation
between the $s$- and $t$-channel diagrams
of Fig.~\ref{f1} would not have been possible.

\subsection{Proof of equivalence to Siegel gauge amplitudes}
\label{ssig}

Let us denote by
\be \label{ex1}
\overline{\PP} = {b_0\over L_0}\, ,
\ee
the propagator in the Siegel gauge.
Then the propagator in a general linear $b$-gauge is given by
\be \label{ex2}
\PP = \overline{\PP} + [\,Q, \,\Omega\,]\, .
\ee
where\footnote{In
 the left definition of \refb{eas9.1}
we could replace $\overline{\cal P}  = b_0/L_0$ by any
other operator $\wt B / \wt L$ where $\wt B$ is an
appropriate linear combination
of the $b_n$'s and $\wt L=\{Q, \wt B\}$, but we have chosen
it to be $b_0/L_0$ to simplify our formul\ae.}
\be \label{eas9.1}
\Omega\equiv   \overline{\cal P}\,\Delta\PP\, ,
\qquad \Delta \PP \equiv \PP - \overline\PP\, .
\ee
Indeed,  recalling that $\{ Q , \PP \} = \{ Q , \overline \PP \} = 1$
we find
\begin{equation}
 [\,Q, \,\Omega\,] = \{ \,Q, \overline\PP\,\}\,
 \Delta \PP -
 \overline\PP \,
 \{ Q , \Delta \PP \, \} =  \Delta \PP \,.
\end{equation}

While equation~(\ref{ex2}) holds for a large class of linear b-gauges,
it can break down when certain conditions
on the operators ${\cal B}_{(g)}$
 are not fulfilled. We will determine these conditions in
\S\ref{s4} and \S\ref{s5}.
We shall now show that assuming relation~(\ref{ex2}) we
can replace all the propagators $\PP$ by the Siegel gauge
propagator $\overline{\PP}$ in on shell amplitudes.
The proof will use manipulations similar to the ones used
in \S\ref{spure}; however the combinatorics will be somewhat
different.

Let us consider
Feynman diagrams with $k$ external legs and $n$ internal
legs. Since we have only three point vertices,
the number $n$ is the same for all the diagrams contributing to
an amplitude at any given order. The external lines are always
labeled, but we shall also label the internal lines as 1, 2, $\cdots$ $n$.
There are $n!$ ways of doing
this, so we sum over all the $n!$
possibilities and divide each diagram by $n!$. We repeat this for
every Feynman diagram contributing to a given amplitude
at a given order.

\begin{figure}
\centerline{
\epsfig{figure=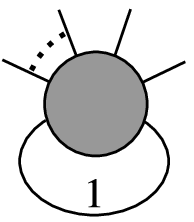, height=3cm}
\hskip 2.5cm
\epsfig{figure=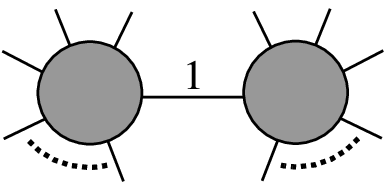, height=3cm}}
 \centerline{(a)\hskip 6.65cm (b) \hskip 1.7cm}
\caption{Irreducible and reducible propagators.} \label{f2}
\end{figure}

Now we collect all Feynman diagrams
contributing to an amplitude  and replace
$\PP$ by  $\overline{\PP}+ Q\Omega - \Omega Q$
in propagator
number 1 in each of these diagrams.
There are two possibilities: (a) the propagator 1
may be irreducible,
\i.e.\
the diagram does not break into two pieces when we cut it,
(b) it may be reducible so that
the diagram breaks into two pieces when we cut it
(see Fig.~\ref{f2}).
We first consider the possibility (a).
For each diagram of this type, we begin with
the $Q\Omega$ term and move the
$Q$ through vertices and propagators  using the relations
\refb{ex3}-\refb{ex4b}.
During the process of moving $Q$ through any of
the other $(n-1)$ propagators,
we again pick up
an extra contribution where the propagator is
replaced by unity. As in \S\ref{spure},
we display this by a collapsed propagator, \i.e\ a
four point vertex, but this time the vertex
 carries the label of the
propagator that collapsed, as  in figure~\ref{f1}.
Other terms where the $Q$ hits
external states vanish
since the external states are BRST invariant.
But this time, because of the irreducibility of propagator 1,
we are left with one more term, $\Omega Q$,  from
bringing the $Q$ back to the
original propagator $1$ on the other side of $\Omega$.
This term cancels the $-\Omega Q$ term of the commutator.
Thus, at the end, the $[Q,\Omega]$ part of
propagator 1 in a given
Feynman diagram of type (a) reduces to
a collection of $(n-1)$
diagrams each of which
has  the operator $\Omega$ on propagator~$1$,
a collapsed propagator in one of the lines
$2$, $\cdots$ $ n \,$, and  propagators
$\PP$ on all other lines.
As before, when we combine the contributions from different
Feynman diagrams of type (a),
there are pairs of
identical
diagrams\footnote{This time we call two diagrams
identical only if they
have both identical topology and
matching labels on the internal
propagators (collapsed or otherwise)
and external lines.}
with collapsed s- and t-channel
propagators.
As collapsed s- and t-channel diagrams differ in sign,
they again cancel pairwise.

The analysis of case (b) is similar,
the only difference
being that $Q$ never comes back to the original
propagator at the
end of the manipulations.   The terms which
arise from manipulating the
$Q\Omega$ part of the commutator
leave behind
diagrams with one collapsed
propagator on one side of the diagram.
The terms which
arise from manipulating the
$-\Omega Q$ part of the commutator
leave behind
diagrams with one collapsed
propagator on the other side of the diagram.
Again, when we combine the contributions
from all the Feynman
diagrams of type (b), the diagrams involving
collapsed propagators
cancel pairwise.

We have thus shown that the commutator term
$[Q, \Omega]$ in propagator 1 in both cases (a) and (b) does
not contribute to the amplitude.
Thus for each original Feynman diagram we are left with
one diagram,
with the Siegel gauge propagator $\overline{\PP}$ on line 1
and the original propagator $\PP$ on all other lines.

We can now
repeat the analysis by replacing propagator 2 in each
diagram by the right hand side of \refb{ex2}.
The only difference
from the previous analysis is that in each diagram
propagator~1
is now the Siegel gauge propagator.
This does not affect our argument, however,
since the Siegel gauge propagator $\overline{\PP}$,
just like $\PP$, satisfies the
 relations \refb{ex4} and~(\ref{ex4b}), \i.e.\
 \be \label{ex5}
  \{Q,\overline{\PP}\}
  = 1\, .
 \ee
Thus at the end of this process we are left
with a sum of diagrams with
propagators 1 and 2 replaced by  Siegel gauge propagators.
Iterating this procedure, we can replace all propagators by Siegel
gauge propagators.

Finally we turn to the external states.
If $\BB_{(1)}$ is a linear
combination of $b_n$'s with $n\ge 0$, as in the case of
Schnabl gauge, then it is possible to choose the cohomology
elements to be the same as the ones used in the Siegel gauge,
\i.e.\ vertex operators of the form $cV$ where $V$ is a dimension
1 matter primary operator.
In general we need to choose
different representatives of the BRST cohomology in Siegel gauge
and a linear $b$-gauge.
However, since we have already
proven decoupling of BRST exact states, we can replace each of the
external states in the linear $b$-gauge by the representative of the
corresponding BRST cohomology class in the Siegel gauge without
changing the amplitude.
This establishes
that all on-shell amplitudes
in a general linear $b$-gauge are the same as those in Siegel
gauge.

\section{Conditions from consistent Schwinger representations
of $1/\LL_{(g)}$} \label{s4}
\setcounter{equation}{0}

The formal manipulations of \S\ref{s2} and \S\ref{s3}
require
that we have a well-defined inverse of the operator
$\LL_{(g)}$
for every $g$,
 in the sense described in \S\ref{s1}.
Indeed $1/\LL_{(g)}$ enters the
expression for the propagator and various manipulations
involving the propagator, -- {\it e.g.} in the proof of
$\{Q,\PP\}=1$.
In this section we shall
investigate under what conditions the
 matrix elements of $1/\LL_{(g)}$
 encountered in the calculation of string field theory amplitudes
can be rigorously defined
 up to terms whose associated Riemann surfaces are localized at the boundary of open string moduli space.

As a warm-up let us recall how
regularization
works in the
Siegel gauge. The propagator $1/L_0$ is defined as
\be \label{eas1269}
{1\over L_0}
\equiv \lim_{\Lambda_0\to\infty}
\int _0^{\Lambda_0}
ds \, e^{-s L_0} \, .
\ee
Using the relation
\be \label{esi0}
L_0 \int_0^{\Lambda_0}
ds \,
e^{-s L_0} = 1 - e^{-\Lambda_0 L_0}\, ,
\ee
 we see that in order for $\int_0^{\Lambda_0}ds\, e^{-s L_0}$ to
 give a proper definition of $1/L_0$ for
 $\Lambda_0\to\infty$, the matrix elements
 of $e^{-\Lambda_0 L_0}$ must vanish in this limit.
Thus we must examine what happens to the
amplitudes when the propagator on a line is replaced by
the operator $e^{-\Lambda_0 L_0}$.  As is familiar,
in the presence of this operator
the Feynman graph line represents a strip of length
$\Lambda_0$ and width~$\pi$.
As $\Lambda_0\to \infty$ the strip becomes infinitely
long and the Riemann surface degenerates. As long as the open
strings propagating along this infinitely long strip carry
positive conformal weight,  this
contribution can be safely
ignored.

Following the same strategy we try
to represent $1/{\cal L}_{(g)}$
as $\int_0^\infty ds \, e^{-s \LL_{(g)}}$, and then
regulate the upper limit of integration over the
Schwinger parameter $s$ using a cutoff $\Lambda_{(g)}$:
\begin{equation} \label{einv0}
\frac{1}{\LL_{(g)}}\, \equiv \lim_{\Lambda_{(g)}\to\infty}
\int_0^{\Lambda_{(g)}} ds
\, e^{-s \LL_{(g)}}\, .
\end{equation}
Now
we have
\be \label{einv1}
\LL_{(g)} \int_0^{\Lambda_{(g)}} ds
\, e^{-s \LL_{(g)}} = 1 - e^{-\Lambda_{(g)}\LL_{(g)}}\, .
\ee
Thus in order that \refb{einv0} gives a proper definition of
$1/\LL_{(g)}$ we need to ensure that in the
$\Lambda_{(g)}\to\infty$ limit the
$e^{-\Lambda_{(g)}\LL_{(g)}}$ term on the right hand side of
\refb{einv1} has vanishing matrix element between any pair
of states which arise in the analysis of the Feynman amplitudes
of string field theory.
Recalling the analysis in the Siegel gauge,
we can easily anticipate that in order to prove the existence
of $1/\LL_{(g)}$, we need to ensure that
insertion of an operator
$e^{-\Lambda_{(g)} {\cal L}_{(g)}}$ produces degenerate Riemann
surfaces in the $\Lambda_{(g)}\to\infty$ limit.

Keeping this in mind, we shall now examine the effect of inserting
an operator of the form $e^{-s \LL_{(g)}}$ into a correlation
function and then study the result in the $s\to\infty$ limit.
For this we shall assume from the beginning that
condition \refb{condreal1},
 $\,\overline{v(\xi)}=v(\bar\xi)$,
is satisfied
 for the vector field $v(\xi)$ associated with ${\cal L}_{(g)}$
so that we can give a
Riemann surface interpretation to the matrix elements of
$e^{-s\LL_{(g)}}$.
We shall find that in the $s\to\infty$ limit
the insertion of $e^{-s\LL_{(g)}}$
produces degenerate surfaces
if the operators ${\cal L}_{(g)}$ also
satisfy condition~\refb{ealt2old}. Thus when these conditions are
satisfied, eq.(\ref{einv0}) gives a proper definition of $1/\LL_{(g)}$.
 In \S\ref{sprop} we will use this result to give a geometric
 interpretation of the propagator $\PP_{(g)}$ for
 regular linear $b$-gauges.
We shall show in \S\ref{s5} that
for these regular
gauge choices our results
in~\S\ref{s3} hold rigorously, \i.e.\ we have
$\{Q,\PP\}=1$ leading to decoupling of pure gauge states
and the correct on-shell amplitudes are produced.

\subsection{Gluing surface states with $e^{-s\LL_{(g)}}$
insertions} \label{sg}

\begin{figure}
\centerline{\epsfig{figure=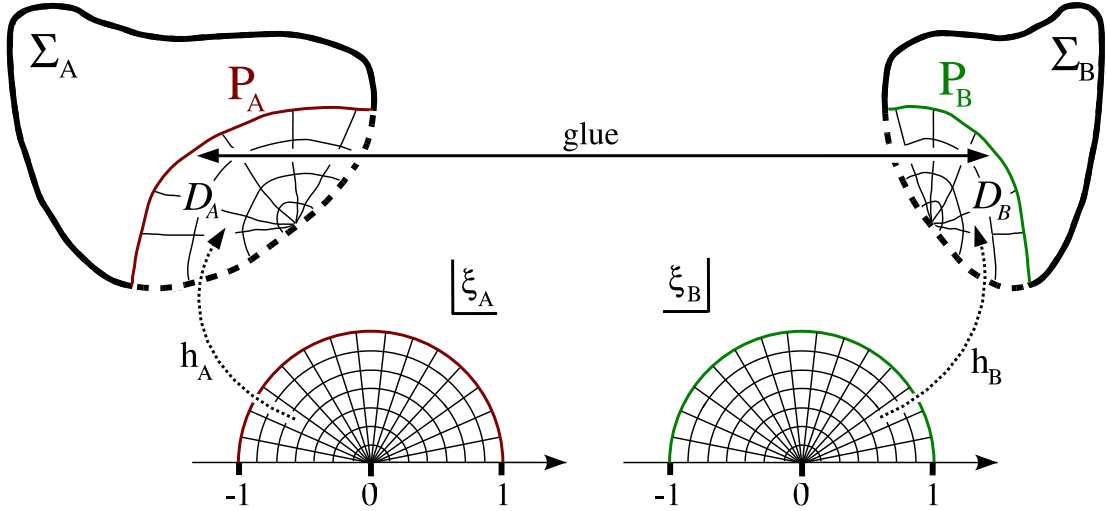, height=7cm}}
\caption{Diagram illustrating gluing of surface states. In this
diagram $P_A$ and $P_B$ denote the  boundaries
created by the
removal of the disks $D_A$ and $D_B$, respectively.
The maps $h_A(\xi_A)$ and $h_B(\xi_B)$ embed the local coordinates
$\xi_A$ and $\xi_B$ into the surfaces $\Sigma_A$ and $\Sigma_B$,
respectively. The gluing of $P_A$ and $P_B$ to form the overlap
$\bra{\Sigma_A} \Sigma_B\rangle$ is induced by the identification
$\xi_A = - \xi_B^{-1}$.
}
\label{fsurface1}
\end{figure}

We want to examine the matrix element of the operator
$e^{-s\LL_{(g)}}$ between two surface states.
The surface states $\bra{\Sigma}$
of interest to us are described by
 a Riemann surface $\Sigma$ with an arbitrary number of boundary components,
with insertions of various vertex operators at the boundary and
integrals of antighost fields, BRST currents, and ghost
number currents
in the bulk. The complete description of $\bra{\Sigma}$
also requires us to specify a marked point
$p$ on the boundary and a map
$h(\xi)$ that takes the
unit half-disk
$|\xi| \leq 1, ~\Im (\xi) \geq 0$
 to a region $\DD$
around $p$ on $\Sigma$,
mapping  $\xi =0$
to $p$, the component of the real axis between
$-1$ and 1 to the component of the boundary of $\DD$ that is
part of the boundary of $\Sigma$, and the unit semicircle
$\xi=e^{i\theta}$ in the upper half-plane to the rest of the
boundary component $P$ of $\DD$.
In that case the state $\bra{\Sigma}$ is defined via the equation
\be
\bra{\Sigma} \phi\rangle = \langle  \OO
\, h\circ \phi(\xi=0)\rangle_\Sigma
\ee
for any Fock space state $\ket{\phi}$.  Here $\langle~\rangle_\Sigma$
denotes correlation function on the Riemann
surface $\Sigma$ and
$\OO$ denotes collectively
all insertions of external vertex operators and integrals of antighost,
BRST and ghost number currents in $\Sigma$.
We shall assume that
all the insertions in $\Sigma$ are {\it outside} the disk $\DD$;
this is necessary in order that the surface state $\bra{\Sigma}$ has
a well-defined inner product with other surface states. In particular
given two such surface states $\bra{\Sigma_A}$ and
$\bra{\Sigma_B}$, we compute their BPZ inner product
$\bra{\Sigma_A}\Sigma_B\rangle$ by
removing the disks $\DD_A$ and $\DD_B$ associated with the
two surfaces, and then gluing $\Sigma_A-\DD_A$ with
$\Sigma_B-\DD_B$ along the new boundary components
$P_A$ and $P_B$, -- generated by
the removal of
$\DD_A$ and $\DD_B$ -- via the map
\be \label{egluemap}
\xi_A = -\xi_B^{-1}\, .
\ee
The result is a correlation function of the operators $\OO_A$ and
$\OO_B$ on a new Riemann surface obtained by gluing
$\Sigma_A$ and $\Sigma_B$ by the
procedure described above (see Fig.~\ref{fsurface1}).

We now turn to the expression of interest:
\be\label{constry}
\bra{\Sigma_A } \, e^{-s {\cal L}_{(g)}} \,\ket{\Sigma_B} \,.
\ee
The goal of our analysis is to show that the operator
insertion can be described as the insertion of a strip-like
domain $\RR(s)$ to the Riemann surface that represents the
overlap $\bra{\Sigma_A } \Sigma_B\rangle$.

\subsubsection{The strip domain $\RR(s)$}

Let us denote the vector field associated with
${\cal L}_{(g)}$ by $v(\xi)$. This vector
field generates a flow $f_s(\xi)$ through the
differential equation\footnote{This differential
equation is equivalent to  $\exp (-s v(\xi) \partial_\xi) \xi = f_s(\xi)$.
This relation also yields the so-called Julia equation $v(f_s(\xi)) = v(\xi) \partial_\xi f_s (\xi)$.}
\begin{equation}\label{efseq}
   \frac{d}{ds}\,f_s(\xi)=-v\bigl(f_s(\xi)\bigr) \,,
   \qquad f_{ s=0 }(\xi)=\xi \,.
\end{equation}
We assume  that $v(\xi)$
is analytic in some neighborhood of the unit circle $|\xi|=1$
and satisfies condition
\refb{ealt2old}.
This means that
\be \label{evcond}
v_\perp(\xi) \ge {r}
\quad \hbox{for $|\xi|=1$
for some $r>0$}\, , \qquad v_\perp(\xi)\equiv
 \Re \left( \bar\xi v(\xi) \right)\, .
\ee
Geometrically,
$v_\perp(\xi)$ represents
the radial component of the vector field $v(\xi)$, and
\refb{evcond} states
that $v(\xi)$ is directed
outwards at every point on the unit circle.
This condition,
together with \refb{efseq},
implies that
\begin{equation}
\label{deratbound}
\p_s |f_s(e^{i\theta})|<0\,, \quad  \hbox{at} \quad s=0\,,
\quad
 0\leq\theta<2\pi \,.
\end{equation}
We do not expect the flow $f_s(\xi)$ to be well
defined for all $\xi$ and arbitrarily large $s$;
 if the vector field $v(\xi)$ has poles,
the function $f_s(\xi)$ will in general have branch cuts.
But the analyticity of the vector field
in a neighborhood of the
unit circle together  with \refb{deratbound}
implies that there is some $s_0>0$ such that
$f_s(e^{i\theta})$ is well defined and one to one
for $0\le s < s_0$. Furthermore \refb{deratbound}
shows that $f_s(e^{i\theta})$ is
inside the unit circle
for $s=0^+$.
 Eq.\refb{evcond} guarantees that the flow
  \refb{efseq} is directed inwards
 when $f_s(e^{i\theta})$ is on the unit
 circle. Therefore once $f_s(e^{i\theta})$ is inside the
 unit circle it must stay inside as we increase
 $s$ as long as the flow is non-singular.
Thus we have
\begin{equation}\label{fsInside}
     |f_s(e^{i\theta})|<1 \qquad \text{ for }
     \quad 0< s< s_0\,,
        \quad 0\leq\theta<2\pi \,.
     \end{equation}
Furthermore the reality condition
in \refb{ealt2} together with
\refb{efseq} tells us that $f_s(\xi)$ is
symmetric under reflection about the real axis as well as about
the origin:
\be \label{eneweq}
\overline{f_s(\xi)} = f_s(\bar\xi),
\quad f_s(-\xi) = - f_s(\xi) \qquad \text{ for }
     \quad 0\le s< s_0\, .
\ee
This in particular implies that $f_s(\xi)$ is real for real $\xi$.
Furthermore
since $f_s(e^{i\theta})$ is well defined and one to one for
$0\le s<s_0$, it must
lie in the upper half plane for $0\le\theta\le\pi$.

For $s=0$, \refb{constry}
reduces to $\braket{\Sigma_A}{\Sigma_B}$
and
can be represented geometrically by gluing
the local coordinates on the surfaces
$\Sigma_A$ and $\Sigma_B$ through the
gluing relation $\xi_A\xi_B=-1$.
As discussed earlier, this prescription glues the
curve $P_A$ : $\xi_A=e^{i \theta}$ on
$\Sigma_A$ to the  curve
$P_B$ : $\xi_B=e^{i(\pi- \theta)}$ on $\Sigma_B$. The effect of
 the operator insertion $e^{-s\LL_{(g)}}$ is to deform
the curve $\xi_A=e^{i \theta}$  into
$\xi_A=f_{s}(e^{i \theta})$.
Due to~(\ref{fsInside}), this new  curve lies within the unit disk of the coordinate
$\xi_A$;
it is
now glued with the curve $\xi_B=e^{i(\pi- \theta)}$ by identifying
the parameter $\theta$ labelling
the two curves.
For the correlator $\bra{\Sigma_A} e^{-s\LL_{(g}}
\ket{\Sigma_B}$, the gluing condition
is thus deformed to $f^{-1}_{ s}(\xi_A)\xi_B =-1$,
or equivalently
\begin{equation}\label{eglueq}
\hbox{Gluing condition:} \qquad   \xi_A=f_{s}(-\xi_B^{-1})\,.
\end{equation}
This corresponds to  inserting an extra
strip $\RR(s)$
between the coordinate curves $\xi_A=e^{i\theta}$ and
$\xi_B=e^{i(\pi-\theta)}$
($0\le\theta\le\pi$) on the surfaces.
Indeed, in the $\xi_A$ plane
the region $\RR(s)$  is  bounded
by the curves
\begin{eqnarray} \label{enewdoman}
Q_A: \quad \xi_A&=&~e^{i\theta} , \hskip16pt\qquad
0\le\theta\le\pi \, , \nonumber \\
Q_B: \quad \xi_A &=& f_s(e^{i\theta}), \qquad
0\le\theta\le\pi \, , \nonumber \\
E_1: \quad \xi_A &=& f_{\beta s} (1) , \hskip2pt
\qquad 0\le \beta \le 1\, , \nonumber \\
E_{-1}:\hskip-1pt \quad \xi_A &=& f_{\beta s} (-1) ,
\hskip-6pt\qquad 0\le \beta \le 1\, .
\end{eqnarray}
The boundary component
$P_A$ of $\Sigma_A-\DD_A$  is glued
with the boundary $Q_A$ of $\RR(s)$
and  the boundary component
$P_B$
of $\Sigma_B-\DD_B$ is glued with the boundary
$Q_B$ of $\RR(s)$.
The trajectory of the point $\xi_A =1$ has
been called $E_1$ and the trajectory of the point $\xi_A = -1$
has been called $E_{-1}$.
Due to eq.\refb{eneweq}
both $E_1$ and $E_{-1}$ lie along the real axis.
This information is
shown in Fig.~\ref{fsurface}.

\begin{figure}
\centerline{\epsfig{figure=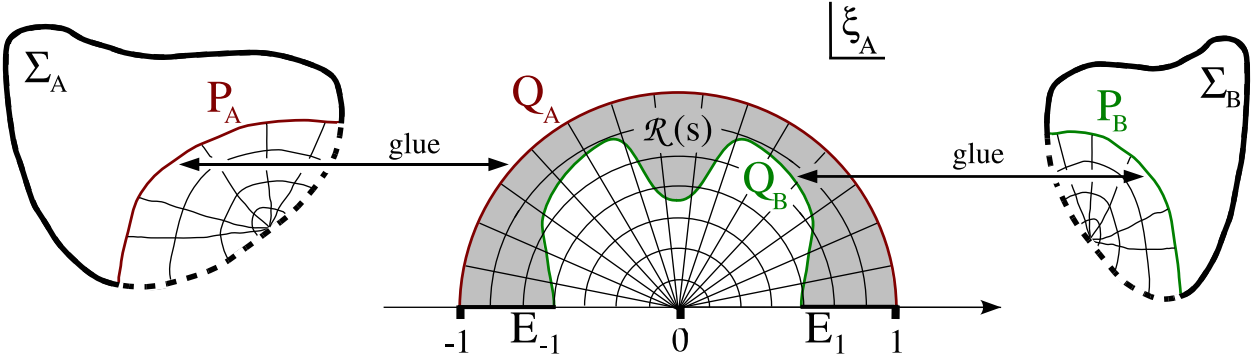, height=5cm}}
\caption{Diagram illustrating the gluing pattern
associated with $\bra{\Sigma_A} e^{-s\LL_{(g)}} \ket{\Sigma_B}$.
The operator insertion effectively
glues the shaded surface $\RR(s)$ to the boundary components
$P_A$ and $P_B$.
The surface $\RR(s)$ is displayed in the $\xi_A$ frame.}
\label{fsurface}
\end{figure}

The gluing relation \refb{eglueq} may be simplified by noting that
the differential equation
\refb{efseq} that determines the function $f_s(\xi)$
is solved by\footnote{Gluing in
  the frame defined by the function $g$
has been discussed
earlier in \cite{0611200}.}
\begin{equation}\label{julia}
   f_s(\xi)= g^{-1}(s+g(\xi)) \,,
\end{equation}
where $g(\xi)$  is a solution to the equation
\begin{equation}\label{egeqn}
     \frac{dg}{d\xi}=-\frac{1}{v(\xi)} \,.
\end{equation}
While the form of $f_s(\xi)$ is not affected by the choice of
integration constant in the solution for $g$, it is convenient
to require $g$ to vanish at $\xi= -1$:
\begin{equation}\label{bcforg}
   g(-1)=0\, .
\end{equation}
Using \refb{julia} the relation \refb{eglueq}
can be expressed as
\begin{equation}\label{juliaglue}
\hbox{Gluing condition:}\quad     g(\xi_A)=s+g(-\xi_B^{-1}) \,.
\end{equation}
This suggests that the piece of surface added by
$e^{-s\LL_{(g)}}$ is most
conveniently represented in a coordinate frame
$w$,
which is
related to $\xi_A$ and $\xi_B$ through the
identifications
\begin{equation}\label{}
    w=g(\xi_A)\,,\qquad
    w=s+g(-\xi_B^{-1})\,.
\end{equation}
These identifications are compatible with the gluing
relation~(\ref{juliaglue}).
Under the map $g(\xi)$,
we obtain a new conformal presentation of the
domain $\RR (s)$ and
of the curves $Q_A, Q_B, E_1,$ and $E_{-1}$
that bound it.
To describe
this we introduce the
curve $\gamma$ describing
the map under $g$ of the half-unit circle:
\be
\label{gamma-def}
\gamma (\theta) \equiv  g (e^{i\theta}) \,, \quad  0\leq \theta \leq\pi\, .
\ee
The curve  $\gamma$ will play a prominent
role in our analysis.
Indeed the curves \refb{enewdoman} in the $\xi_A$ plane
are mapped
by $g$  to
\begin{eqnarray} \label{ewmap}
Q_A: \quad w &=& g(e^{i\theta}) = \gamma(\theta)
\qquad\qquad\qquad\hskip57pt
     \quad  0\leq \theta \leq\pi\, , \nonumber \\
Q_B: \quad w &=& g(f_s(e^{i\theta})) = s + g(e^{i\theta})
= s + \gamma(\theta)
\qquad
     \quad  0\leq \theta \leq\pi\, , \nonumber \\
E_1: \quad
w &=& g(f_{\beta s} (1) ) = \beta s + g(1)
= \beta s +\gamma(0),
     \qquad 0\le \beta \le 1\, , \nonumber \\
E_{-1}: \quad     w &=& g(f_{\beta s} (-1) )
= \beta s + g(-1) = \beta s,
   \hskip17pt   \qquad 0\le \beta \le 1\, .
\end{eqnarray}
We have made repeated use of \refb{julia} at various steps
in \refb{ewmap}.
Thus the domain $\RR(s)$ in the $w$ coordinate system
is
bounded by the curves
$Q_A=\gamma$,
$Q_B=\gamma+s$ and the two horizontal
 line segments $E_1$ and $E_{-1}$
that connect  the
endpoints of the curves $Q_A$ and $Q_B$.
We impose open string boundary conditions
on $E_1$ and $E_{-1}$.
This surface has been
shown schematically in Fig.~\ref{fshape}.
\begin{figure}
\centerline{\epsfig{figure=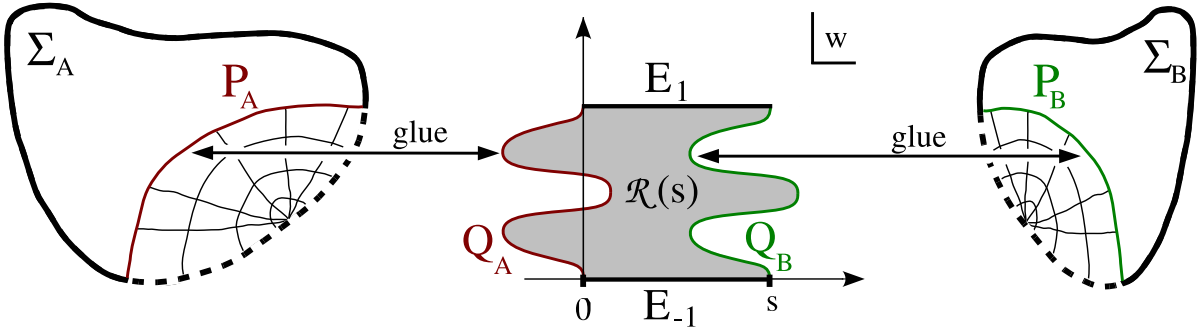, height=4.5cm}}
\caption{
Diagram illustrating $\bra{\Sigma_A}
e^{-s\LL_{(g)}}\ket{\Sigma_B}$
with the surface $\RR(s)$ displayed in the $w$ frame.
The boundaries $Q_A$ and $Q_B$ of
$\RR(s)$ are described by
the curves $w=\gamma(\theta)$ and
$w=s+\gamma(\theta)$, and the horizontal
boundaries correspond to
$\Im(w)=0$ and $\Im(w)=g(1)$.}
\label{fshape}
\end{figure}

So far in our analysis we have restricted $s$ to be in the range
$0\le s<s_0$ (recall ~\refb{fsInside}).  The reason is that the curve
$f_s(e^{i\theta})$ will typically fail to exist for sufficiently large $s$
whenever the vector field $v(\xi)$ has singularities inside the unit disk.
This, we claim, is only a coordinate problem -- the
coordinate frame $\xi_A$
is not suitable to describe sufficiently large deformations.
Instead, as
we will explain, we can use the $w= g(\xi)$ frame that proved
useful above to describe arbitrarily large deformations.

Indeed, to extend our result
 to arbitrary $s>0$
let us first show that if $e^{-s_j\LL_{(g)}}$ is
represented  by the surface ${\cal R}({s_j})$ for $j=1,2$
then
$e^{-s_1\LL_{(g)}}\,e^{-s_2\LL_{(g)}}=
e^{-(s_1+s_2)\LL_{(g)}}$
is represented by the surface
${\cal R}({s_1+s_2})$. In other words, the
surfaces ${\cal R}({s_1})$ and ${\cal R}({s_2})$
 glue nicely to form a longer surface ${\cal R}({s_1+s_2})$.
 This follows immediately
from the fact that the gluing curve $\gamma+s_1$
in the $w_1$ frame
associated with $\RR(s_1)$
is identical,
up to a translation, to the gluing curve $\gamma$
in the $w_2$ frame
associated with $\RR(s_2)$.
Thus
the surfaces join smoothly to form a
longer surface ${\cal R}({s_1+s_2})$
in a frame $w_{12}$ which is related to
$w_1$ and $w_2$ through the simple
identifications $w_{12}=w_1$ and $w_{12}=w_2+s_1$.
Clearly, we can iterate this procedure and build a strip of
arbitrary length $s$ by smoothly joining
short strips of length smaller than $s_0$.
Thus the operator $e^{-s\LL_{(g)}}$
indeed corresponds to the insertion
of the surface $\RR(s)$ even for arbitrarily large $s$.

\subsubsection{Properties of the sewing curve $\gamma$}

We shall now prove some general properties of
the curve $\gamma(\theta) = g(e^{i\theta})$ which
describes the ragged edge
$Q_A$ (and, by translation, $Q_B$) of the region $\RR(s)$.
Integrating eq.\refb{egeqn} along the unit circle $\xi=e^{i\theta}$
and noting that the boundary condition~\refb{bcforg}
means that $\gamma(\pi)=0$, we find
\be \label{einteg}
\gamma(\theta)=g\left(e^{i\theta}\right)
= -\int_\pi^{\theta} \, id\theta' {e^{i\theta'} \over
v\left(e^{i\theta'}\right)}
= \int_\theta^{\pi} d\theta' \, {i\over  u(\theta')}\, ,
\ee
where
\be
\label{defutheta}
u(\theta') = e^{-i\theta'} v\bigl(e^{i\theta'}\bigr)\, \qquad
\hbox{and} \qquad \Re (u(\theta'))
 =v_\perp (e^{i\theta'})
\geq r\,,
\ee
as a consequence of equation \refb{evcond}.
Short calculations then give bounds on the real and imaginary
parts of $i/u(\theta')$:
\be \label{eftbound}
0<\Im \Bigl({i \over  u(\theta')}\Bigr)
\le {1\over r}\,,
\qquad \Bigl|\Re \Bigl({i \over  u(\theta')}\Bigr)\
\Bigr| \le
{1 \over 2 r}\, .
\ee
Equations \refb{einteg} and \refb{eftbound} lead to
several important conclusions. First of all we have
\be \label{emonotone}
\p_\theta\Im
\left( \gamma(\theta)\right) =  -\Im \Bigl({i \over  u(\theta)}\Bigr)
< 0\, ,
\ee
\i.e.\
$\Im \left( \gamma(\theta)\right)$
is a monotonically
decreasing
function of $\theta$.
Thus the curve $\gamma$
never intersects itself.
Moreover,
the surface $\RR(s)$,
 which is swept out by horizontal
 translations of $\gamma$ is
 well defined in the $w$ frame.
Second we have
\be \label{ewidth}
0<
\Im \left(\gamma(0) \right)
=  \int_0^\pi \, d\theta' \,\,\Im  \Bigl({i \over  u(\theta')}\Bigr)
\le {\pi\over r}\, .
\ee
 This together with $\Im (\gamma(\pi)) = 0$ shows that
the region $\RR(s)$ has a finite
and non-vanishing vertical width.
Therefore we can always rescale $v(\xi)$
by a positive real number
to make this width $\pi$:
\be
\label{vert_norm}
\Im (\gamma(0)) = \pi\,.
\ee
 Clearly such a rescaling
affects  neither
the gauge condition $\ointop d\xi v(\xi) b(\xi)
\ket{\psi_{(g)}}=0$
 nor the requirements~(\ref{eonepossible})
 on the coefficients of $v$.
We shall assume from
now on that this has been done, and the  $r$
in the bound  \refb{evcond} refers to the $v(\xi)$ normalized in this
manner. Eq.\refb{ewidth} then gives
\be \label{erbound}
0< r\le 1 \, .
\ee
Finally
it follows from eqs.\refb{einteg} and \refb{eftbound} that
the net horizontal spread $d$ in the curve
$\gamma(\theta)$
is bounded from
above:\footnote{Note that
while the condition on
$v_\perp$ given in \refb{defutheta} is sufficient for getting
a finite vertical width
 and finite horizontal spread,
it may not be necessary.
For example if $v(\xi)$ has isolated zeroes on the unit circle
such that the integral $\int^\theta d\theta' e^{i\theta'}/v(\theta')$
is finite for every $\theta$, we may still be able to get a
curve $\gamma(\theta)$ with all the
desirable properties.
 Though vector fields of this type cannot be analytic
 at
 these isolated zeros, they may still correspond to consistent gauge choices.
 Gauges in which $\BB_{(g)}$ is the zero mode of the antighost
 in the coordinate frame
 of 'wedge states' are of this
 kind. We thank Leonardo Rastelli for discussions on this point.
\label{footgood}}

\be \label{econseq}
d\equiv   \left.\Re\left(\gamma(\theta)
\right) \right|_{max}- \left.\Re\left(\gamma(\theta) \right)
\right|_{min}\le {\pi \over 2 r}\, .
\ee

Additional properties of the curve $\gamma$ arise from use of
the conditions $v(\bar \xi) = \overline{v(\xi)}$ and
$v(-\xi) = - v(\xi)$ in \refb{049irjfgi84}.
Indeed we readily see from the definition
\refb{defutheta} that
\be
\label{upiminustheta}
u(\pi-\theta) = - e^{i\theta} v(-e^{-i\theta}) = e^{i\theta} v(e^{-i\theta})
= \overline{e^{-i\theta} v(e^{i\theta})} = \overline{u(\theta)}\,.
\ee
We can use this to show
that the real part of $\gamma(0)$ vanishes:
\be
\Re (\gamma(0)) = \Re  \int_0^\pi  d\theta'  {i\over u(\theta')}
= {1\over 2} \int_0^\pi   d\theta'
\Bigl( {i\over u(\theta')} - {i\over
\overline{u(
\theta')}} \Bigr)
= {1\over 2} \int_0^\pi   d\theta' \Bigl( {i\over u(\theta')} - {i\over u(\pi-
\theta')} \Bigr) =0\,.
\ee
We therefore conclude that
 \begin{equation}\label{}
\gamma(0) = i\pi\,.  \end{equation}
 It also follows from \refb{upiminustheta} and \refb{einteg} that
 \be
 \overline{\gamma (\theta)} = \int_\theta^\pi  d\theta' {-i\over u(\pi-\theta')}
= \int_{\pi-\theta}^0  d\theta' \,{i\over u(\theta')}
= \int_{\pi-\theta}^\pi  d\theta' \,{i\over u(\theta')}+
 \int_{\pi}^0  d\theta' \,{i\over u(\theta')}\,. \ee
The last integral on the right-hand side is equal to $-\gamma(0) = -i\pi$,
so we get
 \be
 \overline{\gamma (\theta)} =
 \gamma (\pi-\theta) -i\pi  \quad \to \quad
 \overline{\gamma(\theta) - {\textstyle{i{\pi\over 2}}} } =
 \gamma(\pi-\theta)   - {\textstyle{i{\pi\over 2}}}\,.
  \ee
This relation implies that the curve $\gamma$ is reflection symmetric
 about the horizontal line
through $w =i\pi/2$ that bisects the strip $\RR(s)$.

We note
that for BPZ invariant
 vector fields  the net horizontal spread
 $d$ defined in~(\ref{econseq})
 actually vanishes.
 Indeed, for a BPZ invariant vector that is also odd under $\xi\to -\xi$
 one has
 \be
 v(\xi) = - \xi^2 v(-1/\xi) = \xi^2 v(1/\xi) \quad \to \quad
 v(e^{i\theta}) = e^{2i\theta} \, v(e^{-i\theta}) \,.
 \ee
 It follows from this and $v(\bar \xi) = \overline{v(\xi)}$
 that $u(\theta)$ is actually real:
 \be
 e^{-i\theta} v(e^{i\theta}) = e^{i\theta} \, v(e^{-i\theta}) =
 \overline{e^{-i\theta} v(e^{i\theta})} \,.
 \ee
Back in \refb{einteg}  we see that
$\gamma(\theta)$ is a curve
along the imaginary axis -- a vertical
line segment from
$i\pi$ to $0$.
 The simplest example of a BPZ even gauge condition is that
 of Siegel gauge, where $v(\xi) = \xi$
 and consequently $u(\theta) =1$
 and  $\gamma(\theta)=i(\pi-\theta)$.
 More general BPZ invariant gauges
 correspond to $\gamma$'s which
 define different parameterizations
 of the vertical segment from $i\pi$ to $0$.
 The surface $\RR(s)$ is  a rectangle for all BPZ invariant gauges
that  satisfy~(\ref{ealt2}).

\begin{figure}
\centerline{
\epsfig{figure=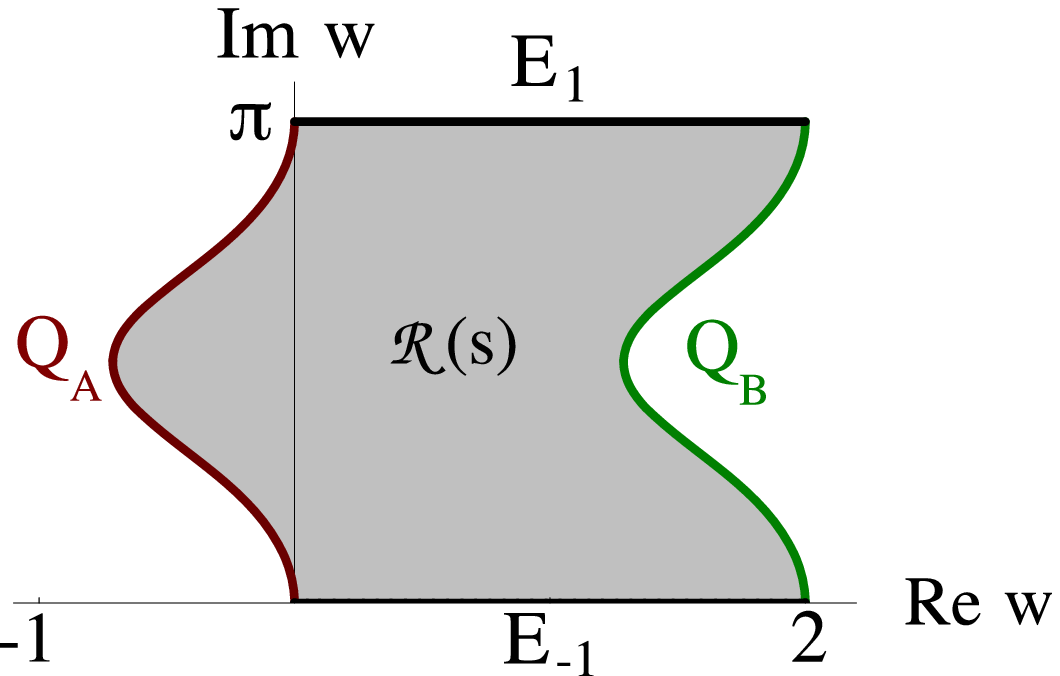, height=5.2cm}
\epsfig{figure=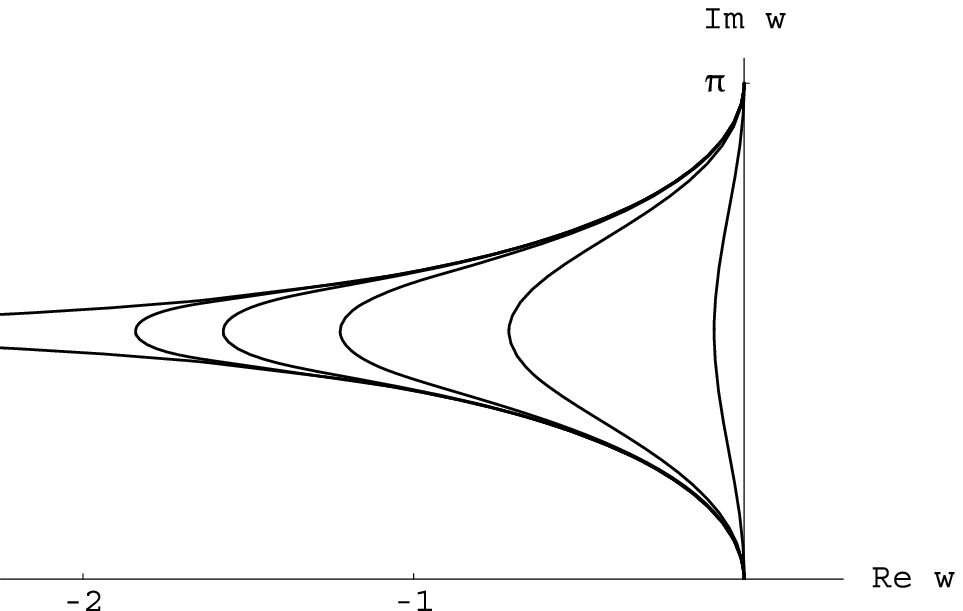, height=5.2cm}
}
\centerline{(a)\hskip 9cm (b)}
\caption{(a) The strip domain
 $\RR(s)$ in the $B^\lambda$ gauge for
 $\lambda=0.1$ and $s=2\,$.
~(b) The curve $\gamma(\theta)$ in the $B^\lambda$ gauge for
$\lambda=1,\,0.1,\,0.01,\,0.001,\,0.0001$, and $0$.
The latter, which corresponds to the Schnabl gauge,
 is singular  at $\theta=\pi/2$. The vertical line along the
 $\Im (w)$ axis corresponds to $\lambda=\infty$, \i.e.\ the
 Siegel gauge.
 }
\label{yyyx}
\end{figure}

\subsubsection{Coordinate frames and examples}

Given a vector field $v(\xi)$ that defines a gauge-fixing
operator $\BB_{(g)}$ by
\be \label{ephys1}
\BB_{(g)} = \ointop {d\xi\over 2\pi i}  v(\xi) b(\xi)\,,
\ee
we
introduce two related coordinate frames.  If we define
$f(\xi)$ via
\be
\label{394ruhfg}
{f(\xi)\over f'(\xi)} = v(\xi)\,,
\ee
then in the
$z= f(\xi)$ frame $\BB_{(g)}$
is the zero mode $b_0$ of
the antighost field. Indeed,
we have
\be \label{ebdef99}
f \circ \BB_{(g)} =  \ointop {d\xi\over 2\pi i}  v(\xi)\,f\circ b(\xi)
= \oint {d\xi\over 2\pi i}  v(\xi)
\left({dz\over d\xi}\right)^2 \,
b(z)=  \oint {dz\over 2\pi i} \,z b(z)\,
=b_0 ~~\hbox{in~}z\hbox{-frame}.
\ee
We also have the $w= g(\xi)$ frame, defined through
\be
\label{ngirt4}
{dg\over d\xi} = -{1\over v(\xi)} \,.
\ee
Perhaps  not surprisingly, $\BB_{(g)}$
(the $g$ subscript is for ghost number
and has nothing to do with the
function $g$)
is the mode $(-b_{-1})$ in the $w$-frame:
\be \label{ebdef69}
g \circ \BB_{(g)} =  \ointop {d\xi\over 2\pi i}  v(\xi)\,g\circ b(\xi)
= \oint {d\xi\over 2\pi i}  v(\xi)
\left({dw\over d\xi}\right)^2 \, b(w)
= - \oint {dw\over 2\pi i} \, b(w)\,
=-b_{-1} ~~\hbox{in~}w\hbox{-frame}.
\ee
Similarly  the  operator $-\LL_{(g)}$ is mapped to
the mode $L_{-1}$ in the $w$-frame -- the
Virasoro mode
associated with translations.
{}From this point of view
it is not surprising that the operator $e^{-s\LL_{(g)}}$ is
represented by a strip of length $s$ in the $w$ coordinate
system.

The relation between $w= g(\xi)$ and $z= f(\xi)$
follows readily from eq.\refb{394ruhfg} and \refb{ngirt4}:
\be
{dg\over d\xi} = - {f'(\xi)\over f(\xi)} \quad \to \quad
g(\xi) = - \ln f(\xi) + \hbox{const.}
\ee
In our conventions $g(-1) = 0$ so we have
\be
\label{gintermsoff}
g(\xi) = - \ln \Bigl[\, {f(\xi) \over f(-1) }\, \Bigr]\,, \qquad
z=f(\xi)=f(-1)e^{-g(\xi)} = f(-1) e^{-w}
\,.
\ee
It is worth noting that, in more generality,
 the operator $\pm\BB_{(g)}$
 ($\pm\LL_{(g)}$)
 is the zero mode $b_0$ ($L_0$)
 in the  coordinate $\tilde z$ related to
  the $\xi$ and $w$ frames through
 \begin{equation}\label{zeromodeframes}
   \tilde z = \tilde z_0 e^{\mp w}=\tilde z_0e^{\mp g(\xi)} \,.
 \end{equation}
for an arbitrary constant $\tilde z_0$.

\medskip
We conclude this subsection with some examples.
For the
$B^\lambda$ gauges the function
$f(\xi)$ associated with the vector $v^\lambda(\xi)$
 is given by $f^\lambda(\xi)$ of
eq.\refb{functionlambda}.  Using \refb{gintermsoff}  we thus have
\begin{equation}
\label{functionlambdagggg}
g(\xi)
= - \ln\,  \Bigl[ { \tan^{-1} (e^{-\lambda} \xi)\over \tan^{-1} (-e^{-\lambda})} \Bigr] \,.
\end{equation}
The left and right boundaries
$Q_A$ and $Q_B$
 of $\RR(s)$ are obtained as
the plot of $g(e^{i\theta})$
 and $s + g( e^{i\theta})$
for $0\le\theta\le\pi$.
These plots are
shown in Fig.~\ref{yyyx}.
We also show the curve $\gamma(\theta)$
for various values of the $\lambda$ parameter.  As we
can see, the horizontal spread of the curve $\gamma(\theta)$ increases
as $\lambda$ decreases.
In particular for $\lambda=0$, \i.e.\
for Schnabl gauge,
$\Re(g(i))=-\infty$,
and the horizontal spread is infinite.
This shows that the strip
domain $\RR(s)$ becomes singular
 in the $w$ frame
in this limit.

\smallskip
It is instructive to consider the BPZ even gauge-fixing operators
 \begin{equation}\label{BBslambda}
     \BB_{(g)}=B^\lambda+(B^\lambda)^\star\,.
 \end{equation}
Since the vector $v^\lambda$ satisfies
all constraints for a regular
gauge, so does the
dual vector $(v^\lambda)^\star$ and,
by linearity, the sum $v^\lambda +  (v^\lambda)^\star$.
This means that the BPZ even
gauges~(\ref{BBslambda})
are regular gauges
 for $\lambda>0$.
As explained before, the BPZ invariance implies
that the horizontal spread  of the curve $\gamma$
vanishes for all $\lambda>0$.
For $\lambda=0$,
$1/u(\theta)$ is proportional to $1/|\theta -{\pi\over 2}|$
near $\theta=\pi/2$. As a result
$\Im(\gamma(\theta))$
computed from \refb{einteg}
diverges logarithmically as $\theta\to \pi/2$.
Thus  the curve
 $\gamma(\theta)$ is again singular and
the width of the strip $\RR(s)$ diverges.
 A strip of divergent width cannot be normalized to width
 $\pi$ by a finite rescaling of the gauge condition~(\ref{BBslambda}).
 On the other hand, a normalization to width $\pi$ is
 possible for all $\lambda>0$, in which case
 the curve $\gamma$ is simply the vertical line
 segment from $i\pi$ to $0$, independent of $\lambda$.
 If we take the $\lambda\to 0$ limit of the curve $\gamma(\theta)$
 with this normalization,
 $\gamma$ approaches the singular
 parametrization
 given by
 \begin{equation}\label{}
 \begin{split}
    \gamma(\theta)&=i\pi\qquad\text{for }
    \quad0\leq\theta<\frac{\pi}{2}\,,\\
    \gamma(\theta)&=0~\qquad\text{for }
    \quad\frac{\pi}{2}<\theta\leq\pi \,.
 \end{split}
 \end{equation}
 Thus we are again lead to  the conclusion
 that the geometric interpretation of the
 gauge condition~(\ref{BBslambda})
 breaks down in the limit $\lambda\to0$.

\subsection{Degeneration and the $s\to \infty$ limit}

Using the general results we have obtained
concerning the region $\RR(s)$
 we can achieve our main goal,
\i.e.\ to show
that in the limit $s\to\infty$ the Riemann surface associated
with the matrix element $\bra{\Sigma_A} e^{-s \LL_{(g)}}
\ket{\Sigma_B}$ is a degenerate Riemann
surface as long as \refb{ealt2} holds.
However for this we need to recall
some facts about degeneration of Riemann surfaces.

Consider a pair of Riemann surfaces $\Sigma_1$ and $\Sigma_2$
with boundaries and
a pair of local coordinates $\eta_1$ and $\eta_2$ around
boundary punctures $p_1\in \Sigma_1$ and $p_2\in \Sigma_2$.
As usual, the coordinates $\eta_i$, $i=1,2$ are restricted to
the canonical upper-half disks $|\eta_i| \leq 1,~\Im (\eta_i) \geq 0$,
and the coordinate maps  take the
boundary $\Im ( \eta_i) =0$ of the half disk to the boundary of
$\Sigma_i$
around the puncture
$p_i$.
The discussion that
follows applies without significant
modification to the case when both
punctures lie on a single Riemann surface
as long as the images of
the unit upper-half disks
$|\eta_i| \leq 1,~\Im (\eta_i) \geq 0$ do not overlap,
so we will continue to
focus on the case when we have two surfaces.

We can sew together the surfaces $\Sigma_1$ and $\Sigma_2$
with a sewing parameter $t \in \mathbb{R}$:
\be
\label{sewrel1}
\eta_1 \, \eta_2 = - t  \,, \quad  0 < t \leq 1\,.
\ee
As usual, this sewing can be done by removing from $\Sigma_1$
and $\Sigma_2$ the images of the half disks $|\eta_i| \leq \sqrt{t}$
and gluing the newly created boundaries.
The sewn surface $\Sigma(t)$
is said to approach degeneration as $t\to 0$.
Degenerations that arise
from sewing are called {\em stable}
degenerations. Sewing also provides a compactification of
the moduli space $t\in (0, 1]$ by the inclusion of the
boundary point  provided by the nodal surface $\Sigma(t=0)$.

\medskip
Having defined degeneration precisely we now want
to show that the composite surface $\Sigma_{AB} (s)$ built in
the previous subsection by gluing the strip $\RR(s)$ to the surfaces
$\Sigma_A$ and $\Sigma_B$ approaches degeneration in the
limit $s\to \infty$.  The strategy is straightforward:  we
introduce two surfaces
$\widetilde\Sigma_A$ and $\widetilde\Sigma_B$
with local coordinates $\eta_A$ and $\eta_B$
 such that the composite surface $\Sigma_{AB} (s)$ arises by sewing
 $\eta_A \, \eta_B = -t$ with some suitable value of $t$ that depends
 on~$s$.  Moreover, $s\to \infty$
must imply  $t\to 0$.

\begin{figure}
\centerline{\epsfig{figure=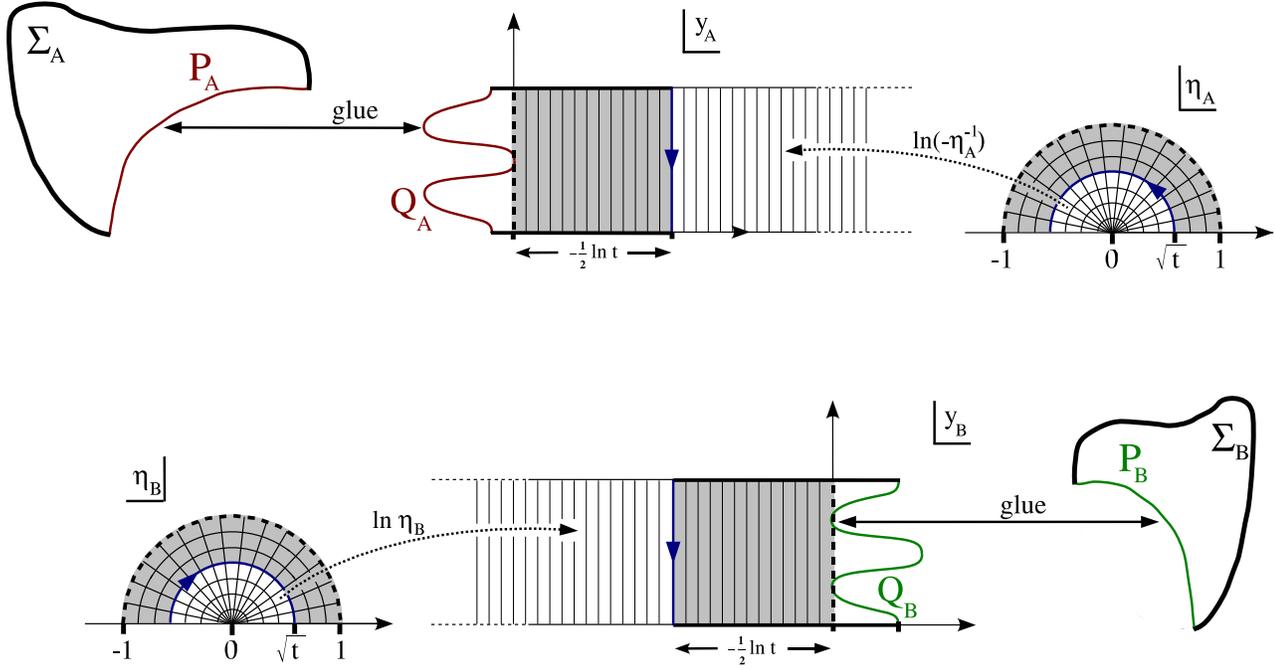, height=9cm}}
\caption{Diagrams illustrating the proof of degeneration
of the surface $\RR(s)$ in the limit $s\to\infty$.
}
\label{figdeg}
\end{figure}

\smallskip
The surface $\widetilde \Sigma_A$ is defined by
gluing the edge
$Q_A$ of  the strip $\RR(s)$ to $P_A$, as before,
but now letting the strip become of infinite
length (see Fig.~\ref{figdeg}).
This introduces
a puncture at the infinite end of the strip.
On the strip we mark a
dotted vertical line immediately to the right of the
ragged curve $Q_A$.  The coordinate
$\eta_A$ around the puncture is defined by
the canonical map that
takes the semi-infinite strip to the right of the
dotted vertical line
to the half-disk $|\eta_A| \leq 1, \, \Im (\eta_A)
\geq 0$. If the strip  is described
with a $y_A$
coordinate in which the dotted line goes from $y_A=0$ to $y_A= i\pi$,
the map is
$\eta_A=- e^{-y_A}$ or $y_A = \ln(-\eta_A^{-1})$.
 The surface $\widetilde \Sigma_B$
is defined analogously: we glue the edge $Q_B$ of $\RR(s)$ to
$P_B$, as before, and let the strip become of infinite length.  This
time $\eta_B$ is defined by the
canonical map from the semi-infinite
strip to the left of the dotted line to the
upper half disk (Fig.~\ref{figdeg}),
\i.e.\ $\eta_B=e^{y_B}$ and thus $y_B = \ln(\eta_B)$.
Note that the
coordinates $y_A$ and $y_B$ differ
from the $w$ coordinate introduced earlier by a simple shift.

Consider now the sewing of $\widetilde
\Sigma_A$ and $\widetilde \Sigma_B$
with
\be \label{esewing}
\eta_A \eta_B = -t
 \,, \quad  0 < t \leq 1
\,.
\ee
It can be performed using cutting curves
$|\eta_A | = |\eta_B| = \sqrt{t}$
and proceeds as follows.  The curve $|\eta_A|= \sqrt{t}$ corresponds
to a vertical line
on the $\widetilde\Sigma_A$
strip a distance
$-{1\over 2} \ln t$ to the right of the dotted
vertical line
 in the $y_A$ frame.
 We amputate
the surface at this line.  Similarly, the curve $|\eta_B|= \sqrt{t}$
corresponds to a vertical line
on the
$\widetilde\Sigma_B$ strip a distance
$-{1\over 2} \ln t$ to the left of the dotted
vertical line
 in the $y_B$ frame.
We amputate this
surface at this line.
 The gluing of these two amputated surfaces is natural,
 because the gluing relation~(\ref{esewing}) in
 terms of the
 coordinates
 $y_A$ and $y_B$ takes
 the simple form $y_A=y_B-\ln t$.
 We thus obtain
a surface in which the distance between the dotted vertical lines is
$-\ln t$.
This surface is,
in fact, the composite surface $\Sigma_{AB} (s)$
built by
gluing the strip $\RR(s)$ to the surfaces $\Sigma_A$ and
$\Sigma_B$
with a value of $s$
 given by
\be \label{estrel}
s = -\ln t + d\,,
\ee
where $d$ denotes the horizontal spread of the curve $\gamma$,
as defined in~\refb{econseq}.
This represents
our composite surface $\Sigma_{AB} (s)$ (with $s\geq d$)
as the result of
sewing
two
auxiliary surfaces $\wt\Sigma_A$
and $\wt\Sigma_B$ via eq.\refb{esewing}. Furthermore we
see from \refb{estrel}
that the limit $s\to \infty$ corresponds to $t\to 0$,
and hence $\Sigma_{AB}(s)$ approaches
degeneration as $s\to \infty$.
This proves the desired result.

\medskip
We have thus shown that, as long as \refb{ealt2} holds,
the Riemann surface associated
with the matrix element $\bra{\Sigma_A} e^{-s \LL_{(g)}}
\ket{\Sigma_B}$ degenerates in the limit $s\to \infty$.
We can also consider matrix elements with products of
multiple operators $e^{-s_i{\cal L}_{(g_i)}}$.
Again, if \refb{ealt2} holds,
\be\label{constry3}
\bra{\Sigma_A } \,\prod_i e^{-s_i
{\cal L}_{(g_i)}}\,\ket{\Sigma_B} \,
\ee
with $s_i\geq0$
represents a degenerate surface  if any of the $s_i\to\infty$.
It is clear that the product can also contain an
arbitrary number of factors $e^{-s_0 L_0}$, because
the vector field
$v(\xi)=\xi$
associated with $L_0$
satisfies~\refb{ealt2}.
Furthermore it should be noted that
this argument is independent of what operators $\OO_A$ and
$\OO_B$ are inserted on the Riemann surfaces $\Sigma_A$
and $\Sigma_B$ since
these do not affect the moduli
of the surface.\footnote{Under certain circumstances insertions
of BRST operators could make the
integrand a total derivative, and
hence, if we wish, we can express the result in terms of
conformal field theory correlation functions
at the boundaries of the region of integration. However if the
whole region of integration is pushed towards the degeneration
limit, the boundaries of the region of integration also reach
the degeneration limit.}
Finally the result quoted above also holds if we insert local
operators (or line integrals of local operators) in between
the $e^{-s_i{\cal L}_{(g_i)}}$ operators in \refb{constry3}.

\medskip
One of the most interesting properties of Siegel gauge is
that amplitudes exhibit off-shell factorization.
We can use the above construction to understand why
this property is so hard to attain and, apparently, occurs
only in Siegel gauge.  Geometrically, the general linear
$b$-gauge propagators add strips of the
form $\RR(s)$.\footnote{The full propagator
 inserts two strips corresponding to $e^{-s\LL_{(g)}}$ for
 two different ghost numbers $g$,
as explained in \S\ref{sprop}, but it seems to us that
 taking this into account
cannot fix the geometrical obstructions
to off-shell factorization
 described below.
}
When the strips become infinitely long the amplitude will
factorize.  We showed above that the insertion of a strip
$\RR(s)$ to the surfaces $\Sigma_A$ and $\Sigma_B$
 with local coordinates $\xi_A$ and $\xi_B$
can be viewed as standard sewing of the surfaces
$\wt\Sigma_A$ and $\wt\Sigma_B$,
 using their local coordinates $\eta_A$ and $\eta_B$.
When the strip becomes infinitely long, the factorization
occurs with off-shell ingredients the surfaces
$\wt\Sigma_A$ and $\wt\Sigma_B$.
On the other hand,
the lower-order off-shell amplitudes in this theory are defined
by the original surfaces $\Sigma_A$ and $\Sigma_B$.  Off-shell
factorization thus requires the conformal identity of
$\Sigma_A$ and $\wt\Sigma_A$ as well as the conformal
identity of $\Sigma_B$ and $\wt \Sigma_B$.  In particular,
this requires that the local coordinates $\xi_A$ and $\eta_A$ be
the same.
But this requirement determines the gauge completely -- only Siegel
gauge satisfies this condition. To see this, let us recall the coordinate
$y_A$ introduced above. It differs from the $w$ coordinate by
a simple shift: $y_A=w+y_0$. The requirement $\eta_A=\xi_A$ for off-shell
factorization can then be written as
\be
  \eta_A=- e^{-y_A} =- e^{-w-y_0} = - e^{-g(\xi_A)-y_0} = \xi_A \,.
\ee
The boundary condition $g(-1)=0$ implies $y_0=0$ and thus
\begin{equation}\label{}
   g(\xi) = -\ln \xi  + i\pi \,.
\end{equation}
From
\begin{equation}\label{}
   \frac{dg}{d\xi}  =-\frac{1}{v(\xi)}\,
\end{equation}
it then follows that $v(\xi)=\xi$.
So we are led to the conclusion that
Siegel gauge is the only regular linear $b$-gauge which exhibits off-shell
factorization.

\subsection{Schwinger parametrization of the propagator}\label{sprop}

Before concluding this section we shall give a geometric
description of the propagator using the geometric description
of $1/\LL_{(g)}$ developed in this section.
 In order to regulate the linear $b$-gauge propagator $\PP$
 we must regulate the ingredients shown in~\refb{propagatorg}.
We  use
\be \label{eas1069}
{\BB_{(g)}\over \LL_{(g)}} = \int_0^{\Lambda_{(g)}} ds_{(g)}
\, \BB_{(g)} \,  e^{-s_{(g)} \LL_{(g)}}\,  .
\ee
where we have introduced a large cutoff
$\Lambda_{(g)}$
for the Schwinger parameter $s_{(g)}$.
Then the regulated propagator
$\PP_{(g)}$ at ghost number $g$
can be written as
\begin{equation}\label{propreg}
\PP_{(g)} = {\BB_{(g-1)}\over \LL_{(g-1)}} \, Q\,
{\BB_{(g)}\over \LL_{(g)}}
= \BB_{(g-1)} \,
\Biggl[\int_0^{\Lambda_{(g-1)}} ds_{(g-1)}
\int_0^{\Lambda_{(g)}} ds_{(g)}
\,   e^{-s_{(g-1)} \LL_{(g-1)}}
 \,  e^{-s_{(g)} \LL_{(g)}}\Biggr]
 \,Q\, \BB_{(g)}
\,  .
\end{equation}
Geometrically the operators of the type
$e^{-s_{(g)} \LL_{(g)}}$
insert strip-like domains $\RR_{(g)}(s_{(g)})$
as discussed in
\S\ref{sg}.
Thus the operator $e^{-s_{(g-1)}
\LL_{(g-1)}}e^{-s_{(g)} \LL_{(g)}}$
can be viewed as the
insertion of a surface created by the gluing of
$\RR_{(g)}(s_{(g)})$ to  $\RR_{(g-1)}(s_{(g-1)})$.
As long as the vector fields associated with
$\BB_{(g)}$ satisfy
the conditions \refb{ealt2} we can set the
upper limits of integration
$\Lambda_{(g)}$ and $\Lambda_{(g-1)}$ in \refb{propreg}
to infinity without encountering any subtlety.
To complete the propagator~(\ref{propreg}),
antighost and BRST insertions have to be
added to the surface,
and the Schwinger parameters $s_{(g-1)}$
and $s_{(g)}$ have to be integrated over.

A particular simple geometric interpretation
can be given to the propagator of alternating gauge,
\begin{equation}
\begin{split}
    {\cal P} &= \frac{{\cal B}_{(1)}}{{{\cal L}_{(1)}}}
    \, Q \, \frac{{\cal B}_{(1)}^\star}{{\cal L}_{(1)}^\star}
    \,\Pi_+
    + \frac{{\cal B}_{(1)}^\star}{{\cal L}_{(1)}^\star}
    \, Q \, \frac{{\cal B}_{(1)}}{{\cal L}_{(1)}}\,\Pi_- \,.
\end{split}
\end{equation}
In this case we only need two types of surfaces,
namely the surface $\RR(s)$ associated with $\LL_{(1)}$
and the surface $\RR^\star(s^\star)$ associated
with $\LL^\star_{(1)}$.
Denoting the vector field associated with
$\LL_{(1)}$ by $v$,
we can derive the following relation
between the boundary curves $\gamma$
and $\gamma^\star$ of $\RR$ and $\RR^\star$:
\begin{equation}\label{}
    \gamma^\star(\theta)
    =\int_\theta^{\pi} d\theta' \, \frac{i}
    {e^{-i\theta'} v^\star\left(e^{i\theta'}\right)}
    =\int_\theta^{\pi} d\theta' \, \frac{-i}
    {e^{i\theta'} v\left(-e^{-i\theta'}\right)}
    =-\int_\theta^{\pi} d\theta' \, \frac{-i}
    {e^{i\theta'} \overline{v\left(e^{i\theta'}\right)}\,}
    =-\overline{\gamma(\theta)} \,,
\end{equation}
where we used~(\ref{ealt2}) and the definition
of $v^\star$ given in eq.(\ref{defvstar}).
This shows that the curve $\gamma^\star$ is the
reflection of the curve $\gamma$ around the
imaginary axis of the $w$-plane. The propagator
$\PP$ in alternating gauge is built from surfaces obtained
by gluing $\RR^\star(s^\star)$ to $\RR(s)$.
For definiteness, let us focus on the surface
associated with the propagator acting on the
subspace of states of even ghost number $g$.
This requires gluing
the right end of $\RR(s)$ to the left end of $\RR^\star(s^\star)$,
as shown in
figure~\ref{figRRstar}(a).
We notice that the surfaces $\RR$ and $\RR^\star$
glue naturally in the
$w$-frame\footnote{In Schnabl gauge,
the surfaces $\RR$ and $\RR^\star$
glue most naturally
in the "sliver frame" $\tilde{z}=-e^{-w}$
where the edges $Q_A$ and
$Q_B$ become vertical lines
and the geometric interpretation of $e^{-sL}
e^{-s^\star L^\star}$ described
in~\cite{0708.2591} can be recovered.}
if we reflect $\RR^\star$ around the imaginary axis, as depicted
in figure~\ref{figRRstar}(b). This reflection
is not necessary if the curve $\gamma$ parameterizes
a precisely vertical line segment. In this case
the surfaces $\RR$ and $\RR^\star$ have the
same shape, and the geometric interpretation
is simply a longer strip $\RR(s+s^\star)
=\RR^\star(s+s^\star)$.
This, of course, describes BPZ invariant
gauges, in which case the propagator could
have been written more suggestively as
$\PP=\BB_{(1)}/\LL_{(1)}$ to begin with.

\begin{figure}
\centerline{
\epsfig{figure=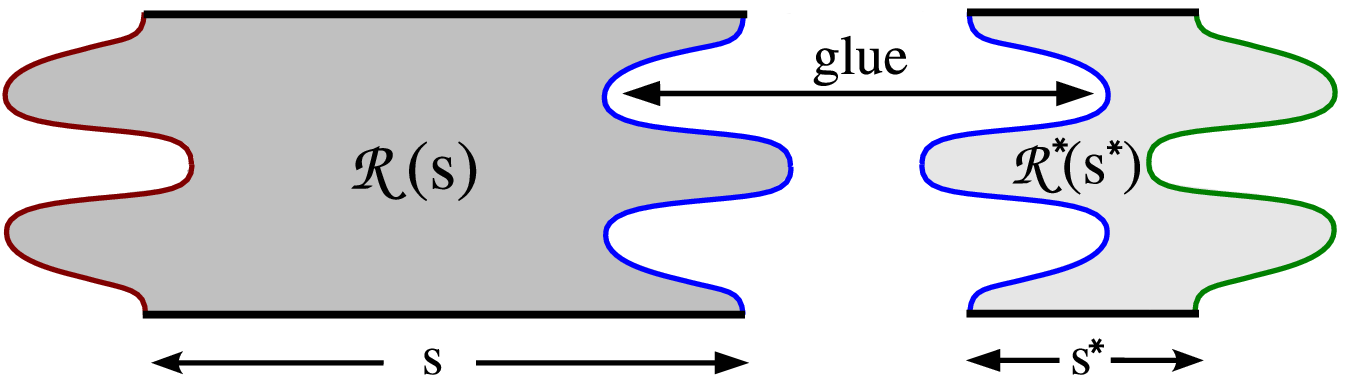, height=2.9cm}
\hskip1.5cm
\epsfig{figure=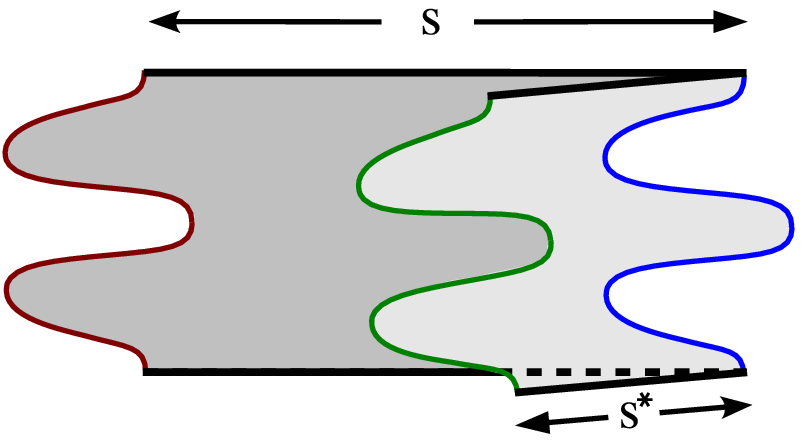, height=3.5cm}
}
\centerline{(a)\hskip 9cm (b)}
\caption{Gluing of surfaces $\RR(s)$ and
$\RR^\star(s^\star)$ for a
geometric
interpretation of the propagator $\PP$ on even ghost
number states in alternating gauge. }
\label{figRRstar}
\end{figure}

\section{On-shell amplitudes revisited} \label{s5}
\setcounter{equation}{0}

In this section we shall use the results of \S\ref{s4} to
show that the formal results of \S\ref{s3} are not affected
by the regularization of the propagator as long as the
 conditions \refb{ealt2} are
satisfied.

\subsection{Decoupling of trivial states}

We shall first
examine the corrections to the relation $\{ Q, \PP\}=1$
which arise when we regulate the propagator, and show,
using the results of  the previous
section,  that these
corrections can be ignored if the
conditions \refb{ealt2} hold.
 The regulated Schwinger parametrization of the operators
 $1/\LL_{(g)}$ introduced above
 in~(\ref{eas1069})
results in the relation
\be \label{eas117069}
\left\{ Q,  {\BB_{(g)}\over \LL_{(g)}}\right\}
= 1 - e^{-\Lambda_{(g)}\LL_{(g)} }\, .
\ee
Use of this identity, eq.\refb{propagatorg},
and $\PP= \sum_g \PP_{(g)} \Pi_g$
 quickly gives
\be
\label{qonPcal}
\{ Q, \PP \} = 1- \sum_g   \left \{ e^{-\Lambda_{(g)}
\LL_{(g)}}
+ e^{-\Lambda_{(g-1)} \LL_{(g-1)}} Q {\BB_{(g)}\over
\LL_{(g)}} + {\BB_{(g)}\over
\LL_{(g)}} Q e^{-\Lambda_{(g+1)}
\LL_{(g+1)}}\right\} \Pi_g\, .
\ee
In the proof of decoupling, $Q$ is moved through the
diagram leaving
factors of one from the commutators with $\PP$.
Those factors represent
collapsed
propagators whose contribution was analyzed in \S\ref{s3}.
In \refb{qonPcal} we have
additional operators
appearing on the right
hand side,
and we need to argue that the contribution from these
additional terms
 vanishes.
Using the  propagator
 \refb{propagatorg}
and the result
of \S\ref{s4} that the insertion of $e^{-s \LL_{(g)}}$ can be
represented geometrically as the insertion of a strip,
we can represent the contribution from a given Feynman diagram
as integrals of appropriate correlation functions
on a  Riemann surface.
As a result in any Feynman diagram
each of these
additional terms discussed above
 is sandwiched
between two surface states
built by the Feynman diagrams.
As we stated when introducing \refb{constry}, the surface states can
carry all kinds of external states, line integrals, or even additional
sewing operations. Our task is to show
that the matrix elements of the additional operators
 on the right-hand side of~(\ref{qonPcal}) vanish
between any
pair of surface states.

The first operator that appears inside the
braces in \refb{qonPcal}
 is exactly of the
type discussed in \refb{constry}, so its
contributions can be ignored
if ${\cal L}_{(g)}$ satisfies the conditions \refb{evcond}.
The second operator is of the form
\be
\int_0^{\Lambda_{(g)}}
 dt \,e^{-\Lambda_{(g-1)} \LL_{(g-1)}}Q {\cal B}_{(g)}
 e^{-t \LL_{(g)}}
 \,.
\ee
This term fits the general structure
described in \S\ref{s4} and
hence   as $\Lambda_{(g-1)}\to
\infty$ we get degenerate surfaces.
Note that this happens for any
non-negative
value of $t$.  If we were to move the BRST
operator to the right of
${\cal B}_{(g)}$ one can get extra terms that
reduce the integral over
$t$ to the endpoints, but even then, those surfaces
are still degenerate.
The third operator within braces in \refb{qonPcal}
is of similar type and
requires no new comments.

All in all, this shows that all the violations of the
$\{Q, \PP\} =1$ identity that arise from
regularization
can be safely ignored
and the decoupling of trivial states will hold.

\subsection{Correct on-shell amplitudes}

Let us now examine in detail
how a regulated
linear $b$-gauge propagator
and a regulated Siegel gauge propagator differ
by $Q$-trivial terms plus
other contributions.  We define
\be \label{eas69}
\Delta\PP \equiv \PP  - \overline\PP\, ,
\ee
as well as
\be\label{eas70}
\Omega \equiv \, \overline\PP\, \Delta\PP\,.
\ee
With unregulated propagators, we would readily find that
$[\,Q, \Omega\,] =  \Delta\PP$, the desired statement that
the difference of propagators is $Q$-trivial.
Using the regulated propagators we now find that
\be\label{eas1171}
[\,Q, \Omega\,] = (1- e^{-\Lambda_0 L_0} )\, \Delta\PP
 - \overline\PP\, \{ Q , \,\Delta\PP\,\}
 \,.
\ee
A short computation using \refb{qonPcal} gives
\be
\{ Q, \Delta\PP\} = \sum_g
\left \{ e^{-\Lambda_0 L_0} - e^{-\Lambda_{(g)}
\LL_{(g)}}
- e^{-\Lambda_{(g-1)} \LL_{(g-1)}} Q {\BB_{(g)}\over
\LL_{(g)}} - {\BB_{(g)}\over
\LL_{(g)}} Q e^{-\Lambda_{(g+1)}
\LL_{(g+1)}}\right\} \Pi_g\, .
\ee
This, together with~\refb{eas1171} now gives
 \be \label{eas149}
(1- e^{-\Lambda_0 L_0} )\Delta\PP =  [Q,\Omega ]
+ \Delta_\Lambda\, ,
\ee
where
\be \label{eas14.1}
\Delta_\Lambda={b_0\over L_0}
\sum_g \left \{ e^{-\Lambda_0 L_0} - e^{-\Lambda_{(g)}
\LL_{(g)}}
- e^{-\Lambda_{(g-1)} \LL_{(g-1)}} Q {\BB_{(g)}\over
\LL_{(g)}} - {\BB_{(g)}\over
\LL_{(g)}} Q e^{-\Lambda_{(g+1)}
\LL_{(g+1)}}\right\} \Pi_g\, .
\ee
This means that we can write \refb{eas149} as
 \be \label{eas151}
\Delta\PP =  [Q,\Omega' ] + \Delta_\Lambda'\, ,
\quad\hbox{with}
\quad
\Omega' = (1- e^{-\Lambda_0 L_0} )^{-1} \Omega, ~~
\Delta_\Lambda' = (1- e^{-\Lambda_0 L_0} )^{-1}\Delta_\Lambda\,.
\ee
We now argue that the terms in
$\Delta_\Lambda'$ give degenerate
surfaces so that their contributions can be ignored.
First consider just
$\Delta_\Lambda$, as given in \refb{eas14.1}.
The operators within
braces give by now  familiar degenerate contributions.
The factor of
$1/L_0$ in front does not change this,
as can be realized by introducing
one more Schwinger parameter to represent this factor.
Finally the factor
 of $(1- e^{-\Lambda_0 L_0} )^{-1}$ which
turns $\Delta_\Lambda$ into $\Delta_\Lambda'$ can be written as
 $\sum_{n=0}^\infty e^{-n\Lambda_0 L_0}$
 and also does not change
 the conclusion.  The key fact in this whole analysis
 is that each operator
 that appears in $\Delta_\Lambda'$
 contains at least one exponential whose argument contains
 a $\Lambda$ parameter that goes to infinity,
 multiplying an admissible
 ${\cal L}_{(g)}$ operator.
 This produces an infinite strip.  Exponentials
 without $\Lambda$ parameters produce
 regular surfaces as long as
 the corresponding vector fields satisfy \refb{ealt2}.
 Once we have an infinite strip, the surface
 is degenerate and its contribution  can be ignored.
This completes our proof that linear $b$-gauges which satisfy the
constraints \refb{ealt2} give the correct
on-shell amplitudes.

We would also like to point out that
the convergence property of amplitudes with $e^{-s\LL_{(g)}}$
insertion for large $s$ guarantees
that the regularization
ambiguities of the kind encountered
in \cite{0708.2591}
are absent for
regular
linear $b$-gauges.
The ambiguous
terms of \cite{0708.2591} contain one or more factor of
$e^{-\Lambda_{(g)} \LL_{(g)}}$ and hence would vanish
in the $\Lambda_{(g)}\to\infty$ limit.
This is of course
consistent with the fact
that regular
linear $b$-gauges reproduce
unambiguously the correct on-shell amplitudes of string theory.

\subsection{Projector gauges} \label{sproj}

Our analysis in the previous sections shows that in gauges
which satisfy the
 conditions
\refb{ealt2} the Feynman amplitudes
of string field theory reproduce correctly the on-shell amplitudes
of open string theory. Unfortunately this argument does not
hold for projector gauges since the insertion of
$e^{-\Lambda_{(g)}\LL_{(g)}}$ into an amplitude
 does not in general
localize the contribution to the boundary of the moduli space
in the limit as $\Lambda_{(g)} \to \infty$.
To demonstrate this, we shall choose the familiar Schnabl gauge  but
a similar analysis can be done for any projector gauge of the type
discussed
below~\refb{slivv}.

\begin{figure}
\centerline{
\epsfig{figure=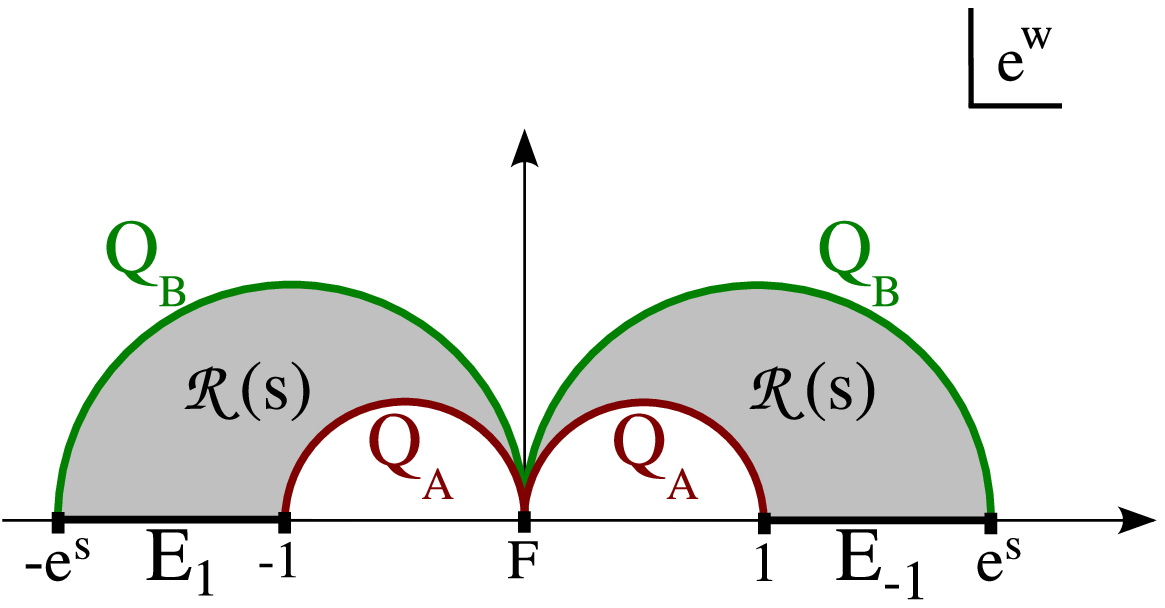, height=4.5cm}
\hskip 0.5cm \epsfig{figure=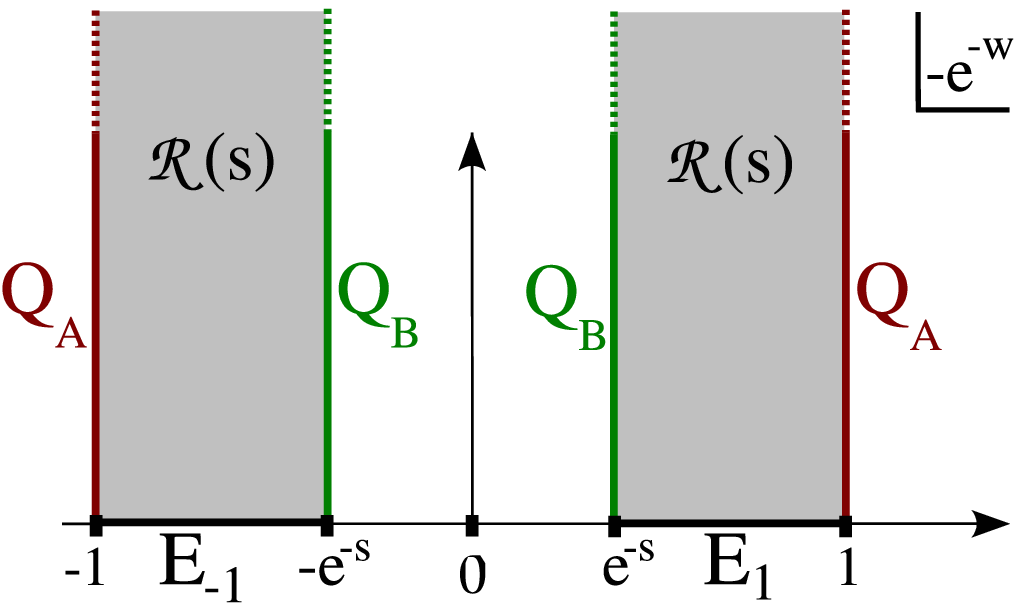, height=4.5cm}}
\centerline{(a)\hskip 8.3cm (b)~~}
\caption{The shape of $\RR(s)$ in Schnabl gauge
displayed in the frame
(a) $e^w$ and  (b) $-e^{-w}$.
In (a) the boundary components $Q_A$ and $Q_B$, which are glued to the
 surface states $\Sigma_A$ and $\Sigma_B$, touch at the fusion point $F$
for all $s$.  In (b) we recover the familiar picture where $Q_A$ and $Q_B$ are
vertical lines and the fusion point is at infinity. }
\label{ewplane}
\end{figure}
As described in Fig.~\ref{yyyx}(b)
the region $\RR(s)$ for
Schnabl gauge looks singular in the $w$ frame since
 the real parts of  the midpoints
($\theta=\pi/2$) of the boundary
components $Q_A$ and $Q_B$
reach $-\infty$. A better understanding of the situation is
obtained by examining this region in the
$\tilde z=e^w$ plane.  This has been
shown in Fig.~\ref{ewplane}(a).\footnote{
For comparison, we also show the strip $\RR(s)$ in
Fig.~\ref{ewplane}(b)
in the frame obtained from $\tilde z$ by the
BPZ map $\tilde z \to - 1/\tilde z$.  The surface
$\RR(s)$ appears as two semi-infinite vertical strips. Up to a constant rescaling,  this frame is the familiar sliver frame.}
As can be seen from this figure, the
midpoints on the boundaries $Q_A$ and $Q_B$ fuse at a single point $F$,
dividing the region $\RR(s)$ into two components.
 This can also be understood as follows.
 The $\tilde z$ coordinate is of the general
 form~(\ref{zeromodeframes}), and thus $-\LL_{(g)}$
 can be interpreted as the generator of rescalings $L_0$ in the
 $\tilde z$ frame. As the fusion point $F$ is the origin of the
 $\tilde z$-frame,
 the action of $\LL_{(g)}$ leaves $F$  invariant and therefore
 $e^{-s\LL_{(g)}}$ fails to separate completely
  the boundary components
 $Q_A$ and $Q_B$ of $\RR(s)$.
   In the computation of $\bra{\Sigma_A}e^{-s\LL_{(g)}}
 \ket{\Sigma_B}$, the boundaries  $Q_A$ and $Q_B$
 are glued to the coordinate curves $P_A$ and $P_B$ of the
Riemann surfaces $\Sigma_A$ and $\Sigma_B$.
The result is a
correlation function on the surface
 $\Sigma_{AB}(s)$
in which the midpoints of the
coordinate curves of $\Sigma_A$ and $\Sigma_B$ remain fused for
all values of $s$.
 Therefore the surface $\Sigma_{AB}(s)$ does not in general exhibit
 open string degeneration in the limit $s\to\infty$.

The simplest example of this phenomenon occurs
in the computation of the four-point function in the
Schnabl gauge:  the matrix element of
$e^{-\Lambda L} e^{-\Lambda^\star L^\star}$
between two three-string
vertices corresponds to the contribution from
a finite point in the moduli space
even for arbitrarily large
$\Lambda$, $\Lambda^\star$ as long as $\Lambda$ and
$\Lambda^\star$ are of the same order~\cite{0708.2591}.
Since $e^{-\Lambda L}$ is the boundary term that arises
in the definition of $1/L$, this shows that $1/L$ is not well defined
acting from the left on states of the form
$\lim_{\Lambda^\star\to\infty}
e^{-\Lambda^\star L^\star}|A*B\rangle$ for a pair of Fock space
states $\ket A$, $\ket B$. Similarly
$1/L^\star$, acting from the right,  is not
well defined on the BPZ conjugate of the above state.
In the case of the four-point function
the problem can be resolved by suitable regularization of the upper
limits of integration, treating $L$ and $L^\star$ symmetrically~\cite{0708.2591}.
It remains to be seen whether the same regularization of the
propagator
produces consistent higher point and/or loop amplitudes as well.

\section{Discussion}

In this paper we studied open string field theory in the class of 
gauges
in which a linear combination $\BB_{(g)}$
of the antighost oscillators
annihilates
the string field of ghost number $g$.
We derived the Feynman rules and showed that
for a wide class of linear $b$-gauges the string field theory amplitudes
reproduce
correctly the on-shell S-matrix elements at the
tree and the loop levels.
Our analysis, however,
does not work for all linear $b$-gauges,
-- certain
regularity 
conditions must be satisfied in order for it
to work. In particular
Schnabl gauge, which has provided
a geometric and algebraic framework to explicitly
construct classical solutions in open string field theory,
fails to satisfy these
 regularity 
conditions.

Schnabl gauge has been known to be subtle
for string perturbation theory for some time.
In particular the analysis of~\cite{0708.2591}  shows that
a consistent off-shell Veneziano
amplitude can be obtained only through a
delicate regularization scheme.
Higher $n$-point tree amplitudes
and loop amplitudes have not been
studied, so there could be additional difficulties  there.
Our analysis shows that these difficulties
can be traced back to the difficulty in defining the
inverse of $\LL_{(g)}\equiv \{Q, \BB_{(g)}\}$ that enters the
definition of the
propagator. The usual  representation
of $1/\LL_{(g)}$ in terms of an integral over a
Schwinger parameter fails due to a non-vanishing boundary
term from the upper limit of integration. This in turn can be traced
back to the fact that
the conformal transformation generated by
$\LL_{(g)}$ in this
gauge does not move the open string midpoint.  As a result, in the
representation as an integral over the Schwinger parameter, the
insertion of
 $1/\LL_{(g)}$ 
does not effectively separate the surfaces it connects even in the
limit where the
Schwinger parameter becomes large.
Notwithstanding these complications
it is still possible that  suitable regularization
 of the propagator 
involving  cut-offs on the Schwinger parameters
and a prescription to take limits
will render
the higher point functions at tree and/or loop level consistent.
This deserves further study.

We constructed a one-parameter
family of regular gauges
which interpolates between Siegel and Schnabl gauge.
It would be interesting to see if this parameter can be
used to regularize Schnabl gauge.  If this is possible,
off-shell amplitudes in Schnabl gauge could be defined by taking the
limit, as we approach Schnabl gauge, of the amplitudes
computed within this family. These results can then be compared
to the off-shell Veneziano amplitudes computed
in~\cite{0708.2591,0609047}
with a different regularization prescription.

Another surprising feature of the Schnabl gauge found
in~\cite{0708.2591} is that the off-shell Veneziano
amplitude does not exhibit off-shell factorization.
We have explained this geometrically and learned that
only in Siegel gauge we expect
off-shell factorization. In other regular $b$-gauges off-shell factorization
is expected to fail because as we attach a propagator to
the coordinate curve
associated with a puncture the natural local coordinate induced by the
strip domain $\RR(s)$ fails to agree with the
original local coordinate.\footnote{We thank
Leonardo Rastelli for raising the question
of off-shell factorization.}  This is, however, not a failure of
the gauge choice. Despite being a desirable feature, off-shell
factorization is not a
requirement we need to impose on a choice of gauge.
The lack of off-shell factorization both for $B^\lambda$ gauges and
for Schnabl gauge is consistent with our proposal to define amplitudes
in Schnabl gauge by taking the $\lambda \to 0$ limit.

Even if open string perturbation theory fails
in Schnabl gauge, it  does not
by itself signal any problem for
the classical solutions
constructed in this gauge since
 they satisfy
the complete set of open string field
theory equations of motion.
Nevertheless, it would be interesting to
obtain exact analytic solutions
in gauges where perturbation theory is well defined, like Siegel gauge.
This will facilitate understanding open string perturbation
theory around the tachyon
vacuum -- in particular open string loop
diagrams which are expected to contain information about closed
string theory.
It will be interesting to see if by making an appropriate gauge
transformation we can convert
Schnabl gauge solutions
into solutions in the family of regular
gauges interpolating between the
Schnabl gauge and the Siegel gauge. This analysis may be
facilitated by the
existence of a continuous
family of gauges: we can now look for infinitesimal gauge
transformations which convert a
solution in one gauge  to another in
a nearby gauge.

\medskip

 Regular linear $b$-gauges satisfy the consistency conditions
 required for a well defined perturbation theory and
our analysis provides an explicit geometric description of the
$1/\LL_{(g)}$ operator. Representing it as an integral of
$e^{-s \LL_{(g)}}$ over the Schwinger parameter $s$ we find that
insertion of $e^{-s\LL_{(g)}}$ into a correlation function
inserts a strip into the Riemann surface on which the correlator
is being computed.
Unlike the case in Siegel gauge, for
which the corresponding
operator $1/L_0$ inserts rectangular strips, here the ends
of the strips which connect to the rest of the Riemann surface are ragged.
For regular linear $b$-gauges, the ends of the strip are parameterized by a
 continuous
curve $\gamma$ that satisfies the following properties:
(i) it is smooth,  (ii) it has finite width,
 (iii) it has finite horizontal spread,
(iv)  it is reflection symmetric about the horizontal line that bisects the strip, (v) it is perpendicular to the open string boundaries, and, (vi) it does not intersect any horizontal line more than once.
These properties followed from our conditions on the vector field
associated with the gauge choice.

 It would be interesting
 to see if consistent linear $b$-gauges arise with weaker conditions.
 In particular,
 any vector field that results in a
 curve $\gamma$ that satisfies the above properties (ii)--(vi)
 but is only continuous as
 opposed to smooth,  may be acceptable.
 This is plausible because the strip domain is still well-defined and reaches open string degeneration for large $s$.
 This class of gauges includes all regular linear $b$-gauges, but also includes gauges in which the vector field
 associated with $\BB_{(g)}$ is not analytic in a neighborhood of the unit circle and may even have zeros on the
 unit circle. Gauges associated with the so-called ``wedge states'' are of this type. The corresponding curves $\gamma$ are continuous but not smooth.

An important difference between a general linear
$b$-gauge and the Siegel gauge is that the propagator in the
former gauge
contains two $1/\LL_{(g)}$
operators (of different ghost numbers) separated by an insertion
of a BRST charge. Thus
in the Riemann surface picture, a propagator will be represented
by a pair of strips separated by the line integral of the
BRST current.
This general structure of the propagator, with two Schwinger
parameters and a BRST insertion in between, suggests that a new
definition of open string amplitudes may be possible.
The usual Polyakov definition of open string amplitudes is closely
related to computations in Siegel gauge, where each Schwinger
parameter is a true modulus of the
Riemann surface.  In linear $b$-gauges
there are two Schwinger parameters and one BRST insertion for each
modulus.   It would be interesting to define, without using
string field theory, string amplitudes of the structure suggested
by perturbation theory in linear $b$-gauges.

\vspace{1cm}

{\bf \large Acknowledgments:}
 We would like to thank
 Ian Ellwood, Theodore Erler, Yuji Okawa, Leonardo Rastelli, and Martin Schnabl
 for discussions and valuable comments on
 a draft version of this paper.
The work of M.K. and
B.Z. is supported in part by the U.S.
DOE grant DE-FC02-94ER40818.
The work of A.S. is supported
in part by the JC Bose Fellowship of DST, India and the
Morningstar Visiting Professorship at MIT. The work
began during the visit of A.S. at the Center for Theoretical
Physics at MIT. A.S. would like to acknowledge warm
hospitality of the members of CTP during the visit.

\end{document}